\documentclass[aps,prb,amsmath,amssymb,superscriptaddress,twocolumn,10ptm]{revtex4-2}

\usepackage{hyperref}
\hypersetup{
    unicode=false,          % non-Latin characters in AcrobatÕs bookmarks
    pdftoolbar=false,        % show AcrobatÕs toolbar?
    pdfmenubar=true,        % show AcrobatÕs menu?
    pdffitwindow=false,     % window fit to page when opened
    pdfstartview={FitH},
    pdfnewwindow=true,      % links in new window
    colorlinks=true,       % false: boxed links; true: colored links
    linkcolor=black,          % color of internal links (change box color with linkbordercolor)
    citecolor=blue,        % color of links to bibliography
    filecolor=magenta,      % color of file links
    urlcolor=blue           % color of external links
}
\usepackage{amsmath}
\usepackage{graphicx}
\usepackage{dcolumn}
\usepackage{physics}
\usepackage[capitalize]{cleveref}
\usepackage{dsfont}
\usepackage{graphicx}
\usepackage{comment}
\usepackage[utf8]{inputenc}
\usepackage{lineno,mathtools,url}
\usepackage{soul}
\usepackage{amssymb,mathrsfs}
\usepackage[dvipsnames]{xcolor}

\newcommand{\pref}[2]{Fig.~\hyperref[#1]{\ref*{#1}(#2)}}

\begin{document}
\title{Many-body theory of false vacuum decay in quantum spin chains
}

\author{Christian Johansen}
\email{christianhoj.johansen@ino.cnr.it}
\affiliation{Pitaevskii BEC Center, CNR-INO and Dipartimento di Fisica, Università di Trento, Trento, Italy}
\author{Alessio Recati}
\email{alessio.recati@cnr.it}
\affiliation{Pitaevskii BEC Center, CNR-INO and Dipartimento di Fisica, Università di Trento, Trento, Italy}
\affiliation{INFN-TIFPA, Trento Institute for Fundamental Physics and Applications, I-38123 Trento, Italy}
\author{Iacopo Carusotto}
\email{iacopo.carusotto@ino.cnr.it}
\affiliation{Pitaevskii BEC Center, CNR-INO and Dipartimento di Fisica, Università di Trento, Trento, Italy}
\affiliation{INFN-TIFPA, Trento Institute for Fundamental Physics and Applications, I-38123 Trento, Italy}
\author{Alberto Biella}
\email{alberto.biella@cnr.it}
\affiliation{Pitaevskii BEC Center, CNR-INO and Dipartimento di Fisica, Università di Trento, Trento, Italy}
\affiliation{INFN-TIFPA, Trento Institute for Fundamental Physics and Applications, I-38123 Trento, Italy}
\date{\today}

\begin{abstract}
In this work we theoretically investigate the false vacuum decay process in a ferromagnetic quantum spin-1/2 chain.
We develop a many-body theory describing the nucleation and the coherent dynamics of true-vacuum bubbles that is analytically tractable and agrees with numerical matrix product state calculations in all parameter regimes up to intermediate times.
This theory allows us to identify different regimes in the parameter space and unravel the underlying physical mechanisms, thus offering new conceptual insight on the microscopic quantum dynamics of metastable states.
In particular, analogies and differences with the cosmological false vacuum decay picture are highlighted and characterized in terms of experimentally observable quantities. 
\end{abstract}

\maketitle

\section{Introduction}

Metastability is a generic physical phenomenon where a system remains trapped in a long-lived excited configuration before transitioning to a stable lower-energy one under the effect of rare events triggered by either thermal or quantum fluctuations.
Such metastable behaviors can be observed in a variety of quantum systems~\cite{Langer1969,Macieszczak2016,Yin2025}, from long-lived $\alpha$-decaying nuclei~\cite{gurney1928wave,krane1991introductory}, to spatially extended condensed-matter systems undergoing a first-order phase transition in both the classical~\cite{debenedetti1996metastable,oxtoby1992homogeneous} and quantum~\cite{Rossini_PR2021} regimes, and cosmological quantum field theories~\cite{Devoto_2022}. 

At low enough temperatures, escape from the metastable state generally requires a quantum tunneling process under the energetic barrier separating it from the true ground state. 
In spatially extended systems, the decay of the metastable state (the false vacuum, FV) generically occurs via the nucleation of ground state (true vacuum, TV) bubbles. 
In recent years, such false vacuum decay (FVD) phenomena have received renewed attention after recent experimental works in ultracold atomic condensates~\cite{Zenesini2024}, quantum annealers~\cite{vodeb2025stirring}, and cold atom-based quantum simulators~\cite{zhu2024probingfalsevacuumdecay}.

In this work, we will investigate FVD phenomena in the specific, yet paradigmatic case of a spin-1/2 chain undergoing a first-order ferromagnetic transition~\cite{Sachdev_2011,Pfeuty1970,Pellissetto_Book}. Such configurations are experimentally realizable on various platforms, including real materials \cite{Exp_TFIM}, well-controlled quantum simulators~\cite{PRXQuantum.2.017003,PRXQuantum.4.027001} based on trapped ions~\cite{de2024observationstringbreakingdynamicsquantum,Luo_2025}, Rydberg atoms in optical tweezers~\cite{keesling2019quantum} and superconducting circuits~\cite{PRXQuantum.5.010317}. Compared to approximate techniques for cold atomic gases~\cite{Drummond:PRXQuantum2021,Billam:PRA2020} the simplicity of the spin-chain model allows numerical simulations of the full quantum dynamics to be performed in a controlled way with matrix-product-state (MPS) techniques~\cite{silvi2019tensor}. 
On the other hand, in spite of recent numerical investigations~\cite{Calabrese2021, Maki2023}, key questions on FVD phenomena are still open and require analytical developments. A theory for the early-time dynamics
was put forward in~\cite{Wilczek2023} and related confinement and meson dynamics were explored in~\cite{Lerose:PRB2020,Surace:NJP2021,Scopa:PRB2022}. Yet, the intermediate timescales where the actual FVD dynamics takes place have remained so far inaccessible to analytical treatment. Note how FVD involves the decay of metastable states under a constant Hamiltonian; as such, it is crucially distinct from those dynamical features that have been predicted for systems slowly driven across a first-order phase transition~\cite{Sinha:PRB2022,Pellissetto_PRB2025,Surace:arXiv2024}.

In this article, we present a comprehensive theoretical study of the range of false vacuum decay effects that can be observed in spin-1/2 chains starting from a metastable initial state. 
Our theoretical approach to the real-time many-body dynamics is distinct from the instanton formalism~\cite{Coleman1977,Devoto_2022}; it is based on a bosonic framework that goes beyond the single-bubble approach of ~\cite{Rutkevich1999} and is closely related to the fermionic squeezed-state approximation introduced in~\cite{Scopa:PRB2022}. Our method is able to accurately reproduce numerical simulations and is applied to understand the physical mechanisms of the bubble-nucleation processes and the following bubble-expansion dynamics. 
Depending on the parameters, several regimes are identified, including one which closely recovers physics similar to that in Coleman's original picture of FVD in the cosmological context~\cite{Coleman1977,Coleman_PRD_II,Linde_1983,Ubaldi_2025}.

The structure of the article is the following. The physical system is introduced in \cref{sec:phys_sys}. The theory for a single bubble is reviewed in \cref{sec:SB}. The bosonic framework that we use to describe the coherent many-bubble dynamics is introduced in \cref{sec:CB}.  After validating the theoretical approach against numerical MPS simulations in \cref{sec:numerics}, we proceed to present our physical results for the two main regimes: \cref{sec:RB} is devoted to the rigid bubble regime, while \cref{sec:XB} deals with the expandable bubble regime where the basic FVD physics is recovered. Conclusions and perspectives are finally drawn in \cref{sec:conclu}. Technical details on the development and the underlying assumptions of the coherent bubble theory are given in \cref{app:CBT} while in \cref{app:ramp} we provide details about the quench protocol adopted. 

\section{Physical system \label{sec:phys_sys}}
We consider the quench dynamics of a spin-1/2 chain described by the Hamiltonian 
\begin{equation}\label{eq:H_quenched}
    H(t)=-J\left[\sum_{n=1}^{N-1}\sigma^x_n\sigma^x_{n+1}+ \sum_{n=1}^N\Big(h_\perp\sigma^z_n+\theta(t)h_\parallel\sigma^x_n\Big)\right]
\end{equation}
where $\sigma_n^\alpha$ are Pauli operators, $h_\perp$ ($h_\parallel$) is the strength of the transverse (longitudinal) field, $J>0$ sets the energy scale and the Heaviside function $\theta(t)$ describes the quench at $t=0$. Throughout this work, we focus on the ferromagnetic phase $h_\perp<1$ in the thermodynamic limit. 

Before the quench, there is no longitudinal field $h_\parallel=0$. Among the two degenerate ground states with opposite magnetization $\left<\sigma^x_n\right>=\pm M=\pm\left(1-h_\perp^2\right)^{1/8}$, we assume that the system is initially prepared in the state with magnetization $+M$, which we refer to as $\ket{0^+}$. Additionally, we shift our energy scale such that $H(t<0)\ket{0^+}=E_{\text{FV}}\ket{0^+}= 0$.
Following the quench, the ground-state degeneracy is lifted by the finite longitudinal field $h_\parallel$; specifically, our choice of $h_\parallel<0$ renders $\ket{0^+}$ metastable.

The spin chain at $h_\parallel=0$ is integrable and diagonalized by fermionic excitations~\cite{Mattis1961,Pikin1966,McCoy1968,Pfeuty1970} described by $\gamma(\theta)$, which annihilates a fermion at momentum $\theta$ and obeys the anti-commutation relations $ \{\gamma(\theta),\gamma^\dagger(\theta')\}=2\pi\delta(\theta-\theta')$.
These fermionic excitations can be associated to smeared-out domain walls of width
$W\approx{\sqrt{h_\perp}}/{(1-h_\perp)}$ \cite{WidthHeyl2020}. 
Their dispersion relation 
\begin{equation}\epsilon(\theta)=2 J\sqrt{1-2h_\perp \cos(\theta)+h_\perp^2}
\end{equation}
displays a finite energy gap $\epsilon(0)=2J(1-h_\perp)$ and a finite bandwidth $\Delta\epsilon=\epsilon(\pi)-\epsilon(0)=4Jh_\perp$.

After the quench, the model becomes non-integrable and $h_\parallel<0$ induces repulsive interactions between fermions of constant strength $\alpha = 2J|h_\parallel| M$~\cite{Rutkevich1999} and couples sectors that differ by an even number of fermions. 
In what follows, we will focus on the $\epsilon(0)/(J |h_\parallel|) \gg 1$ regime where the fermion creation/annihilation processes are weak and can be treated perturbatively in $h_\parallel$.
Note how this perturbative regime is in contrast to the  $\epsilon(0)/(J h_\parallel) \ll 1$ regime where the coupling between different fermion-number sectors is non-perturbative and metastable phenomena are washed out. Our treatment of the perturbative regime is well-known from previous works~\cite{Rutkevich1999} and is explained in detail in \cref{sec:sys}. In a nutshell, this perturbative fermionic theory amounts to rewriting the non-integrable Hamiltonian $H(t>0)$ in \cref{eq:H_quenched}, in terms of the fermions that diagonalize the integrable model $H(t<0)$. The Hamiltonian can then be split into terms that conserve the fermion number and a term that does not. The non-number conserving term can then be treated perturbatively~\cite{Birnkammer2022}.

A key element of the FVD process are the so-called bubbles of TV, separated from the surrounding FV by a pair of fermionic domain walls.
To categorize the nature of the bubbles we introduce the bubble rigidity parameter $K\equiv\alpha/\Delta\epsilon$. 
As we are considering bubbles with vanishing total momentum the inverse rigidity $1/K$ quantifies the maximum spatial expansion $\ell_{\rm max}$ that bubbles can undergo as an amount  $\alpha\ell_{\rm max}$ of repulsive interaction energy between fermions is transferred into the kinetic energy of the fermions that is bounded from above by $2\Delta\epsilon$.

When $K>1$ only a small amount of interaction energy can be transferred into kinetic energy of the fermions, so the bubble has a rigid character and maintains a well-defined size.
In the opposite limit $K\ll1$, the bubble is able to expand under the effect of the repulsive force between fermions. 
In the following we will consider parameter sets $R_{1/2}$ in the rigid bubble (RB) regime with $K\approx 4$ and parameter sets $X_{1/2}$ in the expandable bubble (XB) regime with $K\lesssim 0.04$.

\begin{figure*}[tbp]
\centering
\includegraphics[width=0.99\textwidth]{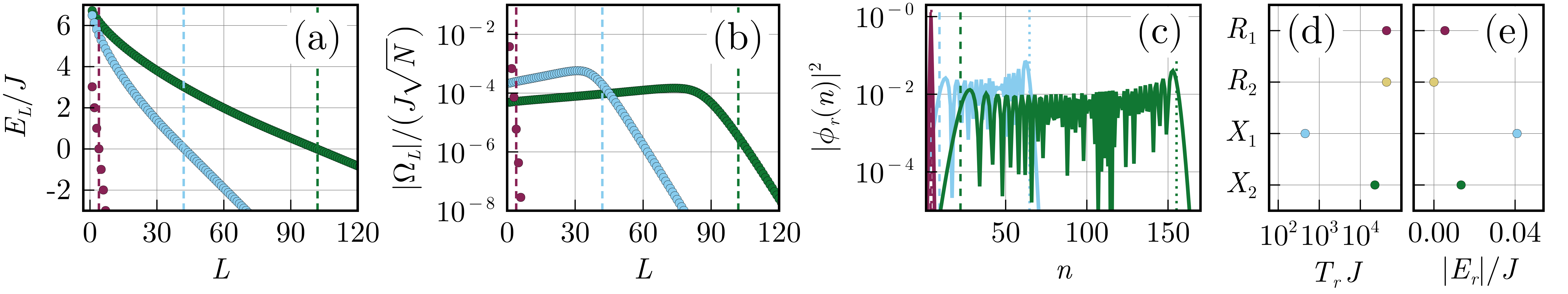}
\caption{
Single-bubble eigenstates for four different parameter sets of $\{h_\perp,h_\parallel\}$. 
The two sets $X_{1}=\{0.75,0.06\}$ and $X_{2}=\{0.75,0.025\}$ are in the expandable bubble (XB) regime, while the two sets $R_{1}=\{0.06, 0.5\}$ and $R_{2}=\{0.06, 0.500675761\}$ are in the rigid bubble (RB) regime, the latter being fine-tuned to have a zero-energy resonant state.
Panel (a) shows the energies $E_L$ of the single-bubble eigenstates; (b) shows their coupling $\Omega_L$ to the initial false vacuum (FV) state $\ket{0^+}$, the resonant eigenstate being indicated by the vertical dashed lines. Panel (c) shows the eigenfunction of the resonant eigenstate. The turning points are plotted with dashed ($l^-_r$) and dotted ($l^+_r$) vertical lines. Panels (d) and (e) give the associated timescale and the energy of the resonant eigenstate, and also serve as the color legend for all subplots. In panels (a-c), results for $R_{1/2}$ are indistinguishable.
}
\label{fig:singleBub}
\end{figure*}

\section{Single-bubble theory \label{sec:SB}}
In both the rigid and the expandable regime one can first restrict to the two-fermion sector, which is equivalent to having only a single bubble present in the entire system. 
As discussed in~\cite{Rutkevich1999,Wilczek2023} and reviewed in \cref{sec:two_fermion}, the two-body problem can be formulated as an eigenvalue problem of the form
\begin{equation}
    \sum_{m\in Z} K_{n-m} \phi_{L,m}-\alpha n \phi_{L,n}=E_L \phi_{L,n},
\end{equation}
where $E_L$ is the energy of the single-bubble state, measured with respect to the energy of the $E_{\text{FV}}$ of the no-bubble state that is set for notational simplicity to zero.  The kinetic energy of the fermions enters the eigenvalue problem through $K_n=\int_{-\pi}^{\pi}\frac{d\theta}{\pi}\epsilon(\theta)e^{i n\theta}$. Diagonalizing the eigenvalue problem yields the energies $E_L$ and the internal wavefunction $\phi_{L,n}$ of the $L$'th zero-momentum two-fermion eigenstate, expressed as
\begin{equation}\label{eq:1bubState}
    \ket{\phi_L}=\frac{1}{\sqrt{2N}}\sum_{n,m}\phi_{L,n}b^\dagger_{m+n}b^\dagger_m\ket{0^+}
\end{equation}
in terms of the real-space fermionic annihilation operators  
\begin{equation}b_m=\int_{-\pi}^\pi \frac{d\theta}{2\pi}\, e^{-i\theta m} \gamma(\theta)\,.\end{equation}
Modulo the domain-wall smearing width $W$, each two-fermion term in the sum on the RHS of \cref{eq:1bubState} describes a single TV bubble of size $n$.
Thus, $\abs{\phi_{L,n}}^2$ represents the weight of a TV bubble of size $n$ in the $L$'th single-bubble eigenstate. The sum over $m$ and the volume factor $1/\sqrt{N}$ account for the delocalization of the center-of-mass over space and make the two-fermion eigenstate \eqref{eq:1bubState} translationally invariant by construction. The quantum number $L$ labeling the eigenstates refers to the relative motion of the two fermions. Note that the spectrum $E_L$ is strictly real, so notions like unstable fluctuation modes with imaginary energies do not arise in our theory.

The energies of the single-bubble eigenstates shown in \pref{fig:singleBub}{a} result from the competition between the energy cost of creating two fermions (i.e., twice the gap) and the energy gained by forming a finite TV domain, associated to the linear repulsive interaction energy between fermions.
A special attention goes to the resonant eigenstate $r = \operatorname*{arg\,min}_L \abs{E_L}$ whose energy $E_r$ is closest to the energy $E_{\text{FV}}=0$ of the initial metastable state. 
As the energy spacing is $\Delta E = 2J\abs{h_\parallel}M$, good resonance $E_r \approx 0$ is naturally found for sufficiently small values of $h_\parallel$, as visible on $X_{1/2}$ in \pref{fig:singleBub}{a}. However, also for relatively large $h_\parallel$, a fine-tuning of $h_\parallel$ can be used to set $E_r\approx 0$ with no appreciable change in the other properties, as illustrated in $R_2$ in \pref{fig:singleBub}{e}.

Most importantly, the rigid vs. expandable nature of the bubbles is apparent in the structure of the eigenfunctions, as shown in \pref{fig:singleBub}{c}.
In the RB regime, a single bubble size dominates each eigenstate and the wavefunction of the closest-to-resonance eigenstate is strongly peaked around the classically expected resonant bubble size in units of the lattice spacing
\begin{equation}l_r^{-}=\frac{2(1-h_\perp)}{\abs{h_\parallel} M}\,.
\end{equation}
In contrast, in the XB regime, many bubble sizes contribute significantly, rendering the eigenstates around resonance ($E_L\sim 0$) spatially delocalized also in the relative distance between the turning points at
\begin{equation}
l^\pm_r=\frac{2(1\pm h_\perp)}{\abs{h_\parallel} M}\,,
\end{equation} marked by vertical lines in \pref{fig:singleBub}{c}.

The couplings $\Omega_L = \bra{0^+}V\ket{\phi_L}$ between the initial FV state and the single-bubble eigenstates are calculated using the perturbative fermionic theory as reviewed in \cref{sec:two_fermion} and, in particular, in  \cref{eq:FVbub_overlap}. Some examples of such calculations are plotted in \pref{fig:singleBub}{b} and exhibit clear qualitative differences between the RB and XB regimes.
Unlike $E_r$, the difference between these couplings is primarily controlled by $h_\perp$, as indicated by the relative difference
\begin{equation}
\label{eq:Omega_decay}
    \left\vert\frac{{\Omega_{r-1}-\Omega_{r+1}}}{{\Omega_r}}\right\vert \approx h_\perp^{-1} - h_\perp,
\end{equation}
which goes to zero as $h_\perp\to1$ in the XB regime, indicating the possibility of coupling also to large bubbles. Furthermore, in contrast to the fast exponential decay of the RB regime, in the XB regime $\Omega_L$ displays a plateau for the positive energy states, which results in a finite number of eigenstates having comparable coupling to $\ket{0^+}$.
Interestingly, the timescale for the resonant processes $T_r = \vert\Omega_r\vert^{-1}$, plotted in \pref{fig:singleBub}{d}, varies by nearly two orders of magnitude across the parameter sets, highlighting the generality of the physical mechanisms.

Based on the single-bubble excitations, one can describe the short-time dynamics arising from creation and annihilation of single TV bubbles as discussed, e.g., in~\cite{Wilczek2023}.
However, it fails to capture the FVD process, which requires accounting for a finite density of TV domains. This is the goal of our coherent bubble theory.

\section{Coherent bubble theory \label{sec:CB}}
As the generic many-body problem is exceedingly complex, we restrict our analysis to a dilute regime where interactions between bubbles can be neglected. Our approach based on a non-interacting bosonic theory can be seen as a bosonic reformulation of the Gaussian approximation based on fermion pairs developed in~\cite{Scopa:PRB2022}. As we will see, this approximation is enough to accurately describe the physics of the FVD process. 

Based on the single-bubble eigenstates introduced above and on their coupling to the initial $\ket{0^+}$ state, we can construct a second-quantized theory in terms of bubble creation operators
\begin{equation}
    d^\dagger_L=\frac{1}{\sqrt{2N}}\sum_{n,m=1}^N\phi_{L,n}b^\dagger_{m+n}b^\dagger_m. 
\end{equation}
Analogously to composite bosons such as bound electron-hole pairs in semiconductor materials, the so-called excitons~\cite{kittel1963quantum,haken1976quantum,Tichy2014}, the bubble operators obey approximately bosonic commutation relations. 
The corrections to the commutation relations are 
quadratic in the fermionic operators [cf. \cref{eq:TI_Commutation}], like the fermion density, meaning that to leading order in density they can be neglected and one can treat $d_L$ as a bosonic annihilation operator.

In this approximation, the time evolution is governed by the effective bosonic Hamiltonian of the form
\begin{equation}
    H_B = \sum_{L} E^{}_L d^\dagger_L d^{}_L + \Omega^{}_L\left(d^{}_L + d^\dagger_L\right).
\end{equation}
Starting from the initial FV $\ket{0^+}$, the time-evolved state can be written as a many-mode coherent state~\cite{Glauber1963} 
\begin{equation}\ket{\psi(t)}=\exp\{\sum_L [\chi^{}_L d_L^\dagger-\chi_L^*d^{}_L]\}\ket{0^+}
\end{equation}
with time-dependent amplitudes straightforwardly given by [cf. \cref{eq:chiAmp} for the details of the derivation]
\begin{equation}\label{eq:timeDep_state}
\chi^{}_L=\frac{\Omega_L}{E_L} \left(e^{-i E_L t} - 1\right)\,.
\end{equation}

Within this Coherent Bubble Theory (CBT), the density of the bosonic bubbles is
\begin{equation}\label{eq:ExcDen}
    \mathcal{N}(t) = \sum_{L} \frac{\langle d^\dagger_L d^{}_L \rangle(t)}{N} =
%    \sum_{L} \frac{2\Omega_L^2}{E_L^2 N} \left[1 - \cos\left(E_L t\right)\right].
    \sum_{L} \frac{4\Omega_L^2}{N} \frac{\sin^2\left(E^{}_L t/2\right)}{E_L^2}.
\end{equation}
Since the coupling $\Omega_L\propto\sqrt{N}$ and decreases exponentially for large $L$ [cf. \cref{eq:FVbub_overlap}], the bubble density $\mathcal{N}$ is indeed independent of the system size $N$.

The bubble density has a further interest as it is related to the FV survival probability $\mathcal{P}(t)$ through
\begin{equation}\label{eq:returnProb}
    \mathcal{P}(t)=\abs{\braket{0^+}{\psi(t)}}^2=e^{-N\mathcal{N}(t)}\,,
\end{equation}
where the exponential function accounts for the overlap of a coherent state with vacuum \cite{Glauber1963}.

The CBT also allows to estimate the density $P_n(t)$ of bubbles of a given size $n$  [cf. \cref{eq:Pn}], from which the TV density is approximately given by
\begin{equation}
    \rho(t) = \sum_{n=1}^N n\, P_n(t),
    \label{eq:rho}
\end{equation}
and the connected spin–spin correlation function can be estimated as [cf. \cref{eq:CorrCBT}]
\begin{equation}\label{eq:corr}
\begin{aligned}
    C_n(t) &= \frac{1}{N} \sum_{m=1}^N \left( \langle \sigma^x_{n+m} \sigma^x_m \rangle - \langle \sigma^x_m \rangle^2 \right) \\
           &\approx 4M^2 \sum_{l=n+1}^N (l - n) P_l(t).
\end{aligned}
\end{equation}
This CBT framework will be our workhorse in the study of the different regimes of FVD that we will address in \cref{sec:RB} and \cref{sec:XB}.

\begin{figure*}[htbp]
\centering
 \includegraphics[width=0.725\linewidth]{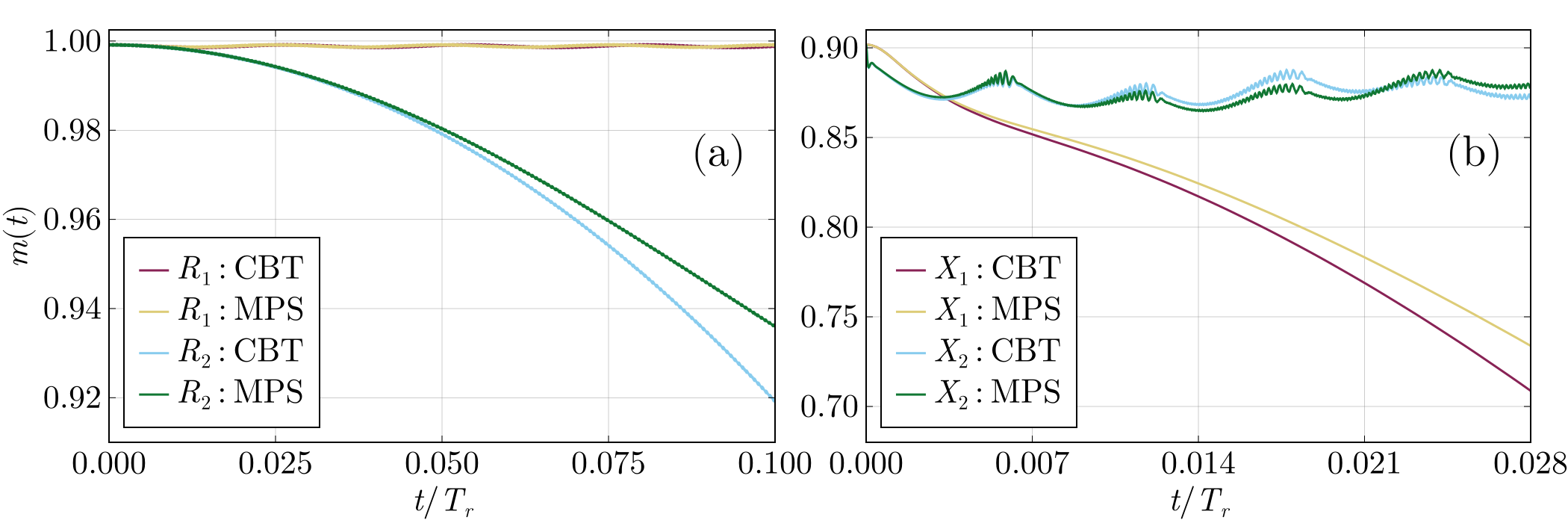}
\includegraphics[width=0.9\textwidth]{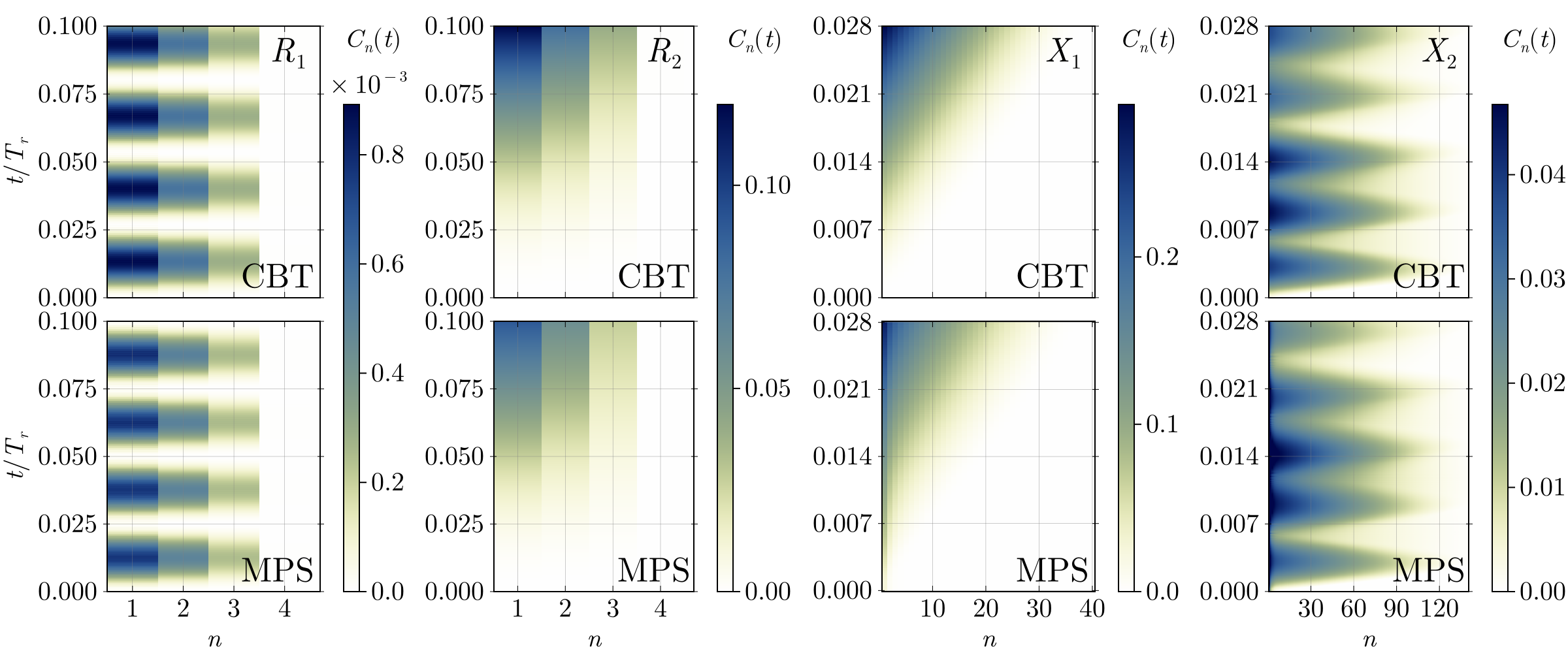}
\caption{Upper panels: comparison between the CBT prediction \cref{eq:magCBT} for the  magnetization and the numerically exact MPS calculation with a bond dimension of 1000. Lower panels: comparison between the spin-spin correlation functions evaluated within the CBT (upper row) with the numerical result computed with matrix-product states (lower row). Same parameters as in the following Figures.
\label{fig:corr_comp}
}
\end{figure*}

\section{Validation of the coherent bubble theory against exact numerical simulations \label{sec:numerics}}

Before delving into these investigations, it is useful to benchmark the predictions of the coherent bubble theory against exact numerical MPS simulations. As we are going to see in this Section, good agreement is found for all parameter sets presented within the time window in which the CBT is expected to be accurate.

The time evolution of the infinite spin-$\tfrac{1}{2}$ chain can be computed numerically using matrix product states (MPS). 
We use the iTEBD algorithm \cite{iTEBD} and consider bond-dimensions of up to 1000, which ensures an excellent convergence for all presented quantities.

To compare the CBT with the MPS results we approximate the magnetization of the spin chain within the CBT (see \cref{app:avg_mag} for details of the derivation)
\begin{equation}
\begin{aligned}\label{eq:magCBT}
      m(t)=M+ \langle G_\Omega\rangle(t)+\langle G_\lambda\rangle(t).
\end{aligned}
\end{equation}
The first contribution arises from the coupling between the false vacuum and one-bubble states
\begin{equation}
\begin{aligned}
    \langle G_\Omega\rangle(t)=\frac{2}{N J\abs{h_\parallel}}\sum_L\frac{\Omega_L^2}{E_L}\left[\cos\left(E_L t\right)-1\right],\\
\end{aligned}
\end{equation}
while the second term is related to the matrix elements of the magnetization operator between  one-bubble eigenstates 
\begin{equation}
\begin{aligned}
    \langle G_\lambda\rangle(t)=\frac{1}{N J\abs{h_\parallel}}\sum_{L,L'}\lambda_{L,L'}\frac{\Omega_L\Omega_{L'}}{E_L E_{L'}}\Big(&\cos\left[(E_L-E_{L'})t\right]\\&-2\cos\left(E_{L}t\right)+1\Big)
\end{aligned}
\end{equation}
with \begin{equation}
\lambda_{L,L'}=-J\abs{h_\parallel} 2M\sum_{n=1}^Nn\phi_{L,n}\phi_{L',n}.
\end{equation}

\begin{figure*}[t]
\centering
\includegraphics[width=0.9\textwidth]{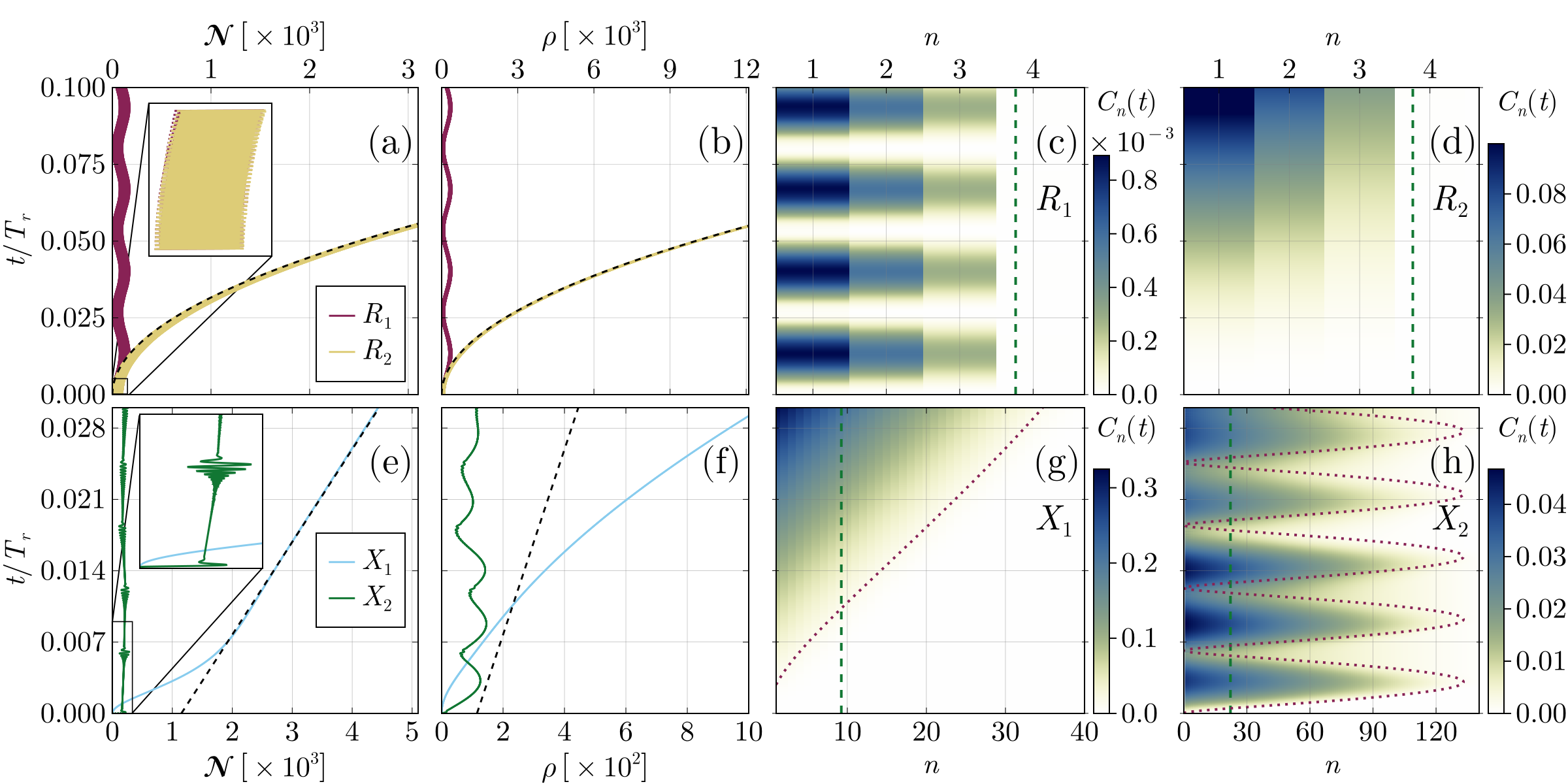}
\caption{Time evolution from the coherent bubble theory for the parameter sets discussed in \cref{fig:singleBub}. (a, e) shows the bubble density $\mathcal{N}(t)$ in \cref{eq:ExcDen}. 
The insets show the short time behavior. The black dashed line in (a) corresponds to the $t^2$ growth predicted by \cref{eq:N_t2_growth}. The black line in (e) corresponds to the linear growth in \cref{eq:N_FGR_adep}, with $a=1/2J$ and $\Omega_{E=0}$ substituted with the average coupling of the two closest-to-resonance states. (b, f) are plots of the approximated TV density $\rho(t)$ in \cref{eq:rho}. The dashed lines in (b,f) correspond to the dashed lines in (a,e) respectively, multiplied by the resonant bubble sizes $l_r$ indicated in (c,g). (c, d, g, h) are the approximated spin-spin correlation functions in \cref{eq:corr}. The dashed vertical lines represent the classical resonant bubble size $l_r$ while the dotted lines show the semi-classical bubble expansion dynamics $r(t)$ of \cref{eq:blochOsc}.
}
\label{fig:evo}
\end{figure*}

The comparison of the approximated magnetization predicted by the CBT with the numerically converged MPS is shown in the upper panels of \cref{fig:corr_comp}.
For both $R_{1}$ and $X_{2}$ the TV density [see \pref{fig:evo}{b,f}] remains small in the considered time window. 
Consequently, the relative error between the MPS and CBT predictions is small and bounded. 
Note how in these  cases the error is not dominated by the emergence of interaction but also influenced by the approximated nature of the magnetization operator used in the CBT, which is considered in the $h_\perp\to0$ limit (see \cref{app:avg_mag} for details).

For $R_2$ and $X_{1}$ the envelope of the TV density grows instead monotonically. 
For long times interactions thus become important and leads to a breakdown of the CBT.
From the comparison of the predictions for the magnetization, we observe that the CBT overestimates the change in magnetization. This suggests that the interactions between bubbles are repulsive and slow down both the expansion and creation of bubbles.
Within the considered time window the maximum relative error of the magnetization $\delta_\text{mag}=\vert m_\text{CBT}(t_\text{max})-m_{\text{TN}}(t_\text{max})\vert/\vert m_{\text{TN}}(t_\text{max})\vert $, is the largest for $X_{1}$ but still remains below $3.5\%$, keeping the CBT prediction close to the converged MPS results for all data  presented in the main text. 

The good agreement seen for the magnetization also holds for the correlation function in \cref{eq:corr}. In the lower panels of \cref{fig:corr_comp} the CBT prediction for the correlation functions are compared with the same correlation functions computed with MPS. 
The agreement is remarkably good considering the simple nature of the approximations used to derive the CBT result presented in \cref{app:CBT}.

The benchmarking against MPS therefore shows that the CBT predictions provide a good quantitative approximation to the full time evolution of the quenched model, in the regime where interactions are not essential.

\section{Results for the rigid bubble regime
\label{sec:RB}}
Having validated the CBT, we now use it to get a deeper understanding of the dynamics of the system.
The results for the RB regime are shown in \pref{fig:evo}{a–d}.
The time evolution of the bubble density $\mathcal{N}(t)$ plotted in \pref{fig:evo}{a} displays signatures of the discreteness of the sum over eigenstates in \cref{eq:ExcDen} and the exponential decay of the coupling $\Omega_L$ for growing $L$ shown in \pref{fig:singleBub}{b}: the strongly coupled high-energy modes give rise to rapid oscillations [see inset in \pref{fig:evo}{a}] superimposed on the slow dynamics governed by a single low-energy near-resonant eigenstate.
As a consequence of the well defined size $l_r$ of the resonant bubble, 
the TV density $\rho$ in \pref{fig:evo}{b} differs from the bubble density $\mathcal{N}$ only by a constant prefactor; the correlation functions in \pref{fig:evo}{c,d} have a rigid envelope that does not extend beyond $l_r$. 

On the other hand, the temporal evolution of the bubble density (and thus the magnitude of the correlation signal) displays a strong dependence on the parameters via the exact value of $E_r$.
For the parameter set $R_{1}$, the near-resonant mode has a small, yet finite energy $E_r \neq 0$ that sets the frequency of the persistent oscillations between the vacuum state $\ket{0^+}$ and the near-resonant eigenstate that are visible in \pref{fig:evo}{a-c}.
With the carefully tuned parameter set $R_2$, we can have $E_r = 0$. In this exactly resonant case, the slow dynamics of the bubble density exhibits a quadratic growth in time, 
\begin{equation}\mathcal{N}_r(t) = \frac{\Omega_r^2 t^2}{N}
\end{equation}
as confirmed by the excellent agreement with the dashed line in \pref{fig:evo}{a}. 

This quadratic growth is reflected in the corresponding increase of the magnitude of the correlations and in a super-exponential decay of the FV survival probability \cref{eq:returnProb}. As already mentioned, the same quadratic law multiplied 
by the resonant bubble size $l_r$  [dashed line in \pref{fig:evo}{b}] properly describes 
the TV density $\rho(t)$ due to the size of the bubbles being constant. In spite of their constant size, the increasing density of the bubbles causes them to start interacting at long times, eventually resulting in a breakdown of the CBT.

\section{Results for the expandable bubble regime}

\label{sec:XB} 

In the XB regime, the single-bubble eigenstates become significantly denser in both energy and coupling strength as seen in \pref{fig:singleBub}{a-b}.
As a result, many such eigenstates now possess comparable energies and couplings, leading to the qualitatively different dynamical behavior shown in \pref{fig:evo}{e–h}.

We first focus on the parameter set $X_{1}$. At short times, the time evolution of the bubble density $\mathcal{N}(t)$, plotted in \pref{fig:evo}{e}, displays a quick transient due to the high-energy eigenstates that get populated during the quench. 
An analog of this fast dynamics with the associated rapid generation of a burst of excitations is also observed when quenching the true vacuum~\cite{Wilczek2023,Scopa:PRB2022}. In \cref{app:ramp} we have verified that, as expected, the amount of such excitations gets suppressed and some form of collapse is observed when the sudden quench is replaced by a slower ramp.

This fast dynamics is followed by a slower evolution characterized by a linear increase of $\mathcal{N}(t)$ in time. This behavior is peculiar to quenches from the false vacuum. It  emerges because the characteristic energy scale over which $\abs{\Omega_L}^2$ appears approximately flat, which corresponds via \cref{eq:Omega_decay} to the timescale \begin{equation}T_\Omega=\frac{1-h_\perp}{J\,|h_\parallel| M}\,,
\end{equation} 
is much shorter than the timescale set by the discreteness of the energy levels, 
\begin{equation}
    T_\Delta = \Delta E^{-1}\,.
\end{equation}
This creates a time window $T_\Omega < t_{\text{FVD}} < T_\Delta$, during which the initial state effectively couples to a continuum of modes, resulting in a Fermi’s Golden Rule (FGR)-type dynamics. Most importantly, note how this FGR dynamics does not require adding any phenomenological interaction-induced broadening of the eigenstates as it was instead done in~\cite{Rutkevich1999}, but naturally stems from the dense nature of the quantum eigenstates.

An immediate consequence of this linear growth of $\mathcal{N}(t)$ is the exponential decay of the FV survival probability $\mathcal{P}(t)$ given in \cref{eq:returnProb}, which is a defining feature of FVD physics.  On top of this, the super-linear growth of the TV density $\rho$ in \pref{fig:evo}{f} signals that the bubbles do expand after being created. This is in agreement with Coleman's original picture~\cite{Coleman1977} and is further highlighted by the stark discrepancy from the straight dashed line in \pref{fig:evo}{f} that assumes a fixed bubble size $l_r$. The bubble expansion is also directly visible in the analogous spatial expansion of the spin correlation function. The bubble expansion speed is determined by the maximum speed at which the fermionic excitations can move, as indicated by the dashed line in \pref{fig:evo}{g}.

The basic FVD dynamics eventually breaks down at long times due to two possible mechanisms.

\subsection{Role of interactions between bubbles}

The first mechanism is the onset of interactions between bubbles: as the density and the size of the bubbles grow, the assumptions underlying the CBT are no longer valid. 
For instance, comparison with numerical MPS calculations for this parameter choice shows that in the FVD time window (up to $\rho\sim 0.1$) the relative error in the magnetization remains at most around $3.5\%$ (cf. \cref{sec:numerics}), but rapidly increases for longer times when interactions between bubbles become crucial.

\begin{figure}
    \centering
    \includegraphics[width=0.75\linewidth]{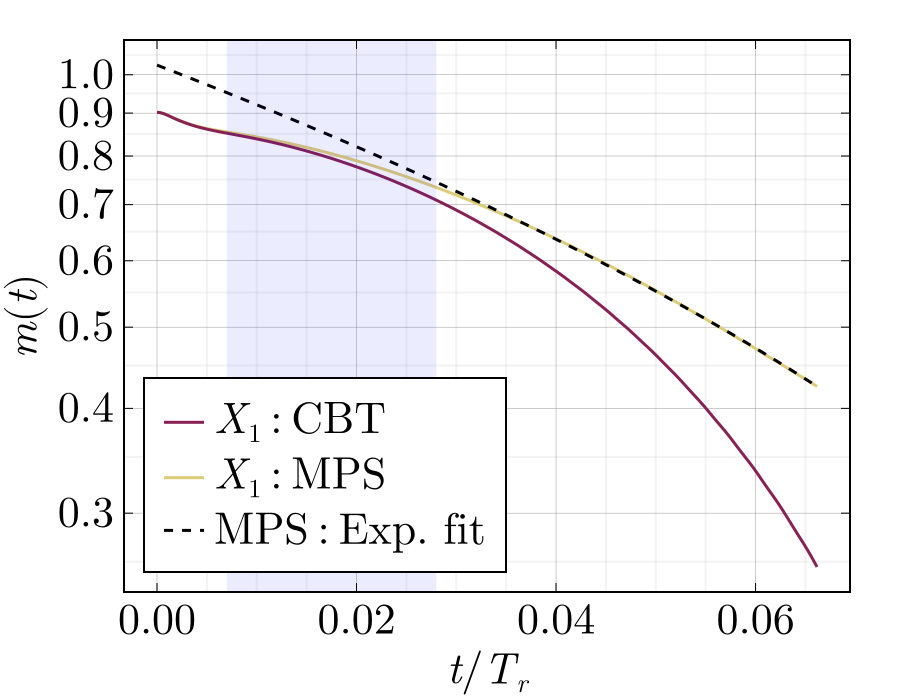}
    \caption{The dynamics of the magnetization for the parameter set $X_1$ at longer times, where the exponential decay reported in \cite{Calabrese2021} becomes visible in the exact MPS calculation, as highlighted by the dashed line showing an exponential fit [the specific fit function is in \cref{eq:m_expFit}]. The blue shaded area indicates the FVD time window where $\mathcal{N}(t)$ grows  linearly [cf. \pref{fig:evo}{e}], and the deviation between CBT and exact MPS remains below 3.5\%.}
    \label{fig:magExp}
\end{figure}

The recent work~\cite{Calabrese2021} has related the FVD dynamics to the exponential decay of the magnetization, which requires an interaction-induced broadening of the two-fermion eigenstates on an intermediate time-scale. In stark contrast to this statement, we find that FVD occurs at much earlier times when interactions are still negligible.
In the FVD time window where bubble nucleation occurs at a constant spatio-temporal rate giving an exponential decay of the FV survival probability in time, the bubble expansion gives a superlinear change of the magnetization but the exponential decay considered in \cite{Calabrese2021} is only found at significantly later times where the bubble density starts being large and the CBT fails.

To confirm this statement, we follow~\cite{Calabrese2021} and we consider the exact numerical MPS calculation of the quantity
\begin{equation}
    F(t)=\frac{m(t)+m(0)}{2m(0)}
\end{equation}
that we attempt to fit with an exponential decay $F(t)\approx b\, e^{-at}$ where $a$ and $b$ are fitting parameters. From this fit, one directly finds an exponential decay of the magnetization
\begin{equation}\label{eq:m_expFit}
    m(t)\approx 2m(0)\, b\, e^{-a t}-m(0)\,.
\end{equation}
While no such behavior is observed within the time window shown in \cref{fig:evo} and in \cref{fig:corr_comp}, this exponential decay clearly emerges as soon as one considers longer times as done in \cref{fig:magExp}. As expected, the exponential behavior in the exact MPS numerics is associated with the failure of the CBT. This indicates that the exponential decay of the magnetization is an intrinsically bubble-interaction-induced effect and is not directly related to the stochastic nucleation of bubbles in Coleman's original picture of FVD~\cite{Coleman1977}.  

\subsection{Effect of the discrete lattice geometry}

The second mechanism is peculiar to the discrete geometry of the chain, as first pointed out in~\cite{TakacsBlochOsc}. Given the periodicity of the fermionic dispersion relation in a lattice, within a semiclassical picture the expansion dynamics of a single isolated bubble displays Bloch oscillations of the bubble size,
\begin{equation}\label{eq:blochOsc}
    r(t)=\frac{1}{J M\abs{h_\parallel}}\left[\epsilon\left(2JM\abs{h_\parallel} t\right)-\epsilon(0)\right],
\end{equation}
whose turning points, as $h_\perp$ gets closer to 1, approximately correspond to the extremes $l_r^\pm$ of the resonant eigenfunction.

This physics is hardly visible in the parameter set $X_{1}$ as the bubble density grows outside the validity range of our CBT well before the onset of Bloch oscillations. 
To probe it, we need to slow down the FVD dynamics compared to the Bloch oscillation one. 
This is done in the parameter set $X_{2}$: by reducing $h_\parallel$, the FVD rate is exponentially suppressed while the Bloch oscillation period increases by a moderate factor around two. 
The resulting physics is illustrated in \pref{fig:evo}{e}: while the early dynamics still features a strong initial increase of the bubble density $\mathcal{N}$ due to the high-energy modes followed by an approximately linear FVD-like behavior (see inset), at longer times the discreteness of the single-bubble eigenstates emerges, making the FGR approximation inaccurate and leading to the Bloch-oscillation-like behavior of the TV density shown in \pref{fig:evo}{f}. 

Even clearer signatures of Bloch oscillations are observed in the correlation function shown in \pref{fig:evo}{h}. Here,
the bubble size predicted by
\cref{eq:blochOsc} is indicated by the dashed line and qualitatively agrees with the edge of the correlation function. After a Bloch period 
\begin{equation}t_{\text{Bloch}}=\frac{2\pi}{2JM\abs{h_\parallel}}\,,
\end{equation} 
the first-generated bubbles have contracted again to almost zero size. 
Of course, as the time progresses the Bloch oscillations of bubbles generated at different times give rise to an overall growth of the background correlation signal which extends up to the maximum bubble size $l_r^+$.
As the Bloch oscillations stem from the discrete lattice geometry, they can be suppressed by reducing the lattice spacing so as to increase the period and the extension of the Bloch oscillations, and thus approach a continuous Ising field theory~\cite{Wilczek2023}.  
 
\section{Conclusions}
\label{sec:conclu}
In this work, we have reported a general theoretical study of the relaxation dynamics of a ferromagnetic spin-1/2 chain in response to a quench of the longitudinal magnetic field. 
Using a bosonic description of the coherent bubble dynamics that is closely connected to the fermionic squeezed-state approximation of~\cite{Scopa:PRB2022}, we have identified very different behaviors depending on the bubble rigidity. 
In particular, a regime displaying an exponential decay of the survival probability of the false vacuum and a quick bubble expansion process is observed. 
In spite of the deep conceptual difference between our theoretical framework and standard semiclassical theories of metastable state decay~\cite{Coleman_PRD_II,Langer1969}, our physical predictions are in full agreement with Coleman's original picture~\cite{Coleman1977} of a stochastic nucleation of isolated bubbles at a given spatio-temporal rate followed by a bubble expansion at relativistic speed. 
Going beyond the literature~\cite{Calabrese2021}, the success of our theory shows that the false vacuum decay process in quantum spin chains does not rely on interactions between the bubbles and does not require including phenomenological broadening terms as it was instead done in~\cite{Rutkevich1999}.
Furthermore, our approach naturally bridges the false vacuum decay process to the Bloch oscillation physics arising from the discrete lattice geometry~\cite{TakacsBlochOsc}.

While our coherent bubble theory is restricted to moderate times when bubbles are too dilute to interact, relaxing this approximation, using for example the cobosonization framework \cite{CobosonizationCombescot}, will provide a tool to explain the late-time exponential relaxation of the magnetization going beyond the rigid bubble description used so far. 
Other interesting directions will be to extend our theory to more complex spin models~\cite{pomponio2025confinement}, in particular spin-1 chains that also include bosonic spin-wave excitation modes. 
This will enrich the model by allowing for spontaneous particle emission processes at the fast-moving bubble interface in connection to white- and black-hole horizon physics~\cite{Wilczek2023}.

An independent study~\cite{Maertens:arXiv2025} with complementary results appeared on arXiv the same day as this submission. 

\acknowledgments

We acknowledge useful discussions with Gianluca Lagnese, Karel Van Acoleyen, Federica Surace and Vyacheslav Rychkov. This work has been supported by the Provincia Autonoma di Trento; by the Q@TN Initiative; by the National Quantum Science and Technology Institute through the PNRR MUR Project under Grant PE0000023-NQSTI, co-funded by the European Union - NextGeneration EU; by the European Union -- NextGeneration EU, within PRIN 2022, PNRR M4C2, Project TANQU 2022FLSPAJ [CUP B53D23005130006]; by European Union’s HORIZON–MSCA-2023-
DN-JD programme under the Horizon Europe (HORIZON) Marie Skłodowska-
Curie Actions, grant agreement 101120296 (TENORS).
Part of the numerical calculations for this work have been performed on the ``Deeplearning cluster" supported by the initiative ``Dipartimenti di Eccellenza 2018-2022 (Legge 232/2016)" funded by the MUR. The data that support the findings of this article are openly available \cite{ZenodoRep}.

\appendix
\onecolumngrid

\section{
Bubble theory}\label{app:CBT}
In this appendix, we derive the CBT starting from the transverse-field Ising model. In \cref{sec:sys} we briefly discuss the mapping of the spin Hamiltonian to a fermion model. Following this, \cref{sec:two_fermion} summarizes the properties of the two-fermion states. These first two sections closely follow the work in \cite{Rutkevich1999}, but are included here to give a self-consistent representation. 
Using the results from the previous sections, we then develop the CBT in \cref{sec:CBT}. Finally, in \cref{sec:scaling} we show how our CBT results scale to a well-defined continuum limit where lattice effects vanish. 
\subsection{Thermodynamic limit and fermion mapping}\label{sec:sys}
We consider a chain with $N$ sites and a spin-$1/2$  at each site, with periodic boundary conditions \cite{IsingBeginners}.
The system's time-evolution is governed by the transverse-field Ising Hamiltonian, where a weak longitudinal field is suddenly switched on at $t=0$
\begin{equation}
    H(t)=-J\left(\sum_{n=1}^{N-1}\sigma^x_n\sigma^x_{n+1}+h_\perp \sum_{n=1}^N\sigma^z_n+\theta(t)h_\parallel\sum_{n=1}^N\sigma^x_n,\right),
\end{equation}
where $\theta(t)$ is the Heaviside function.
Prior to the quench, at $t\leq0$, the system is diagonalized by a Jordan-Wigner transformation followed by a Fourier and a Bogoliubov transformation \cite{IsingBeginners}. 
The diagonalized prequench Hamiltonian takes the form
\begin{equation}\label{eq:H_t0}
H(t<0)=\sum_{\tilde{\theta}}\epsilon_{\tilde{\theta}} \tilde{\gamma}_{\tilde{\theta}}^\dagger\tilde{\gamma}_{\tilde{\theta}},
\end{equation}
where $\tilde{\theta}\in\{\pm \frac{(2n-1)\pi}{N},\,n=1,...,N/2\}$ is the discrete quasi-momentum for a system with an even number of fermions \cite{IsingBeginners}.  

The fermionic annihilation operators $\tilde{\gamma}_{\tilde{\theta}}$ describe excitations with the dispersion relation $\epsilon_{\tilde{\theta}}$ given by
\begin{equation}\label{eq:HTFIM_disp}
    \epsilon_{\tilde{\theta}}=2J \sqrt{1 - 2 h_\perp  \cos\left(\tilde{\theta}\right) + h_\perp^2}\,.
\end{equation}
The annihilation operators obey the fermionic anti-commutation relations
\begin{equation}\label{eq:fermionAC_finite}
    \{\tilde{\gamma}_{\tilde{\theta}},\tilde{\gamma}_{\tilde{\theta}'}\}=\{\tilde{\gamma}_{\tilde{\theta}}^\dagger,\tilde{\gamma}_{\tilde{\theta}'}^\dagger\}=0,\quad \{\tilde{\gamma}_{\tilde{\theta}},\tilde{\gamma}_{\tilde{\theta}'}^\dagger\}=\delta_{\tilde{\theta},\tilde{\theta}'}.
\end{equation} 
The ground state of the prequenched system is two-fold degenerate (for $h_\perp<J$) and leads to two degenerate groundstates $\ket{0_\pm}$ with longitudinal magnetization $\pm M=\pm\left(1-h_\perp^2\right)^{1/8}$.
As mentioned in the main text, we consider quenches from $\ket{0_+}$.

Considering the $N\to \infty$ thermodynamic limit, the momentum spacing, $\Delta_\theta=2\pi/N$, becomes infinitesimal resulting in the following relations
\begin{equation}
    \begin{aligned}
        \delta_{n\%N}=&\lim_{N\to\infty}\frac{1}{N}\sum_{\tilde{\theta}}e^{i\tilde{\theta}n}=\int_{-\pi}^\pi\frac{d\theta}{2\pi}e^{i\theta n},\\
        \lim_{N\to\infty}\frac{\delta_{\tilde{\theta}}}{\Delta_\theta}=&\delta(\theta)=\lim_{N\to\infty}\frac{1}{\Delta_\theta N}\sum_ne^{i\theta n}, \\
        \lim_{N\to\infty}\frac{1}{\Delta_\theta}=&\delta(0).
    \end{aligned}
\end{equation}
For a more streamlined notation, the periodic nature of the discrete Kronecker delta will not be written explicitly.
In the thermodynamic limit, the $\tilde{\gamma}'s$ obey the anti-commutation relations
\begin{equation}
    \{\tilde{\gamma}(\theta),\tilde{\gamma}^\dagger(\theta')\}=\Delta_\theta\delta(\theta-\theta'),
\end{equation}
while the Hamiltonian in \cref{eq:H_t0} takes the form
\begin{equation}
    \lim_{N\to\infty}H(t<0)=\int_{-\pi}^\pi \frac{d\theta}{\Delta_\theta}\epsilon(\theta) \tilde{\gamma}^\dagger(\theta)\tilde{\gamma}(\theta),
\end{equation}
where $\epsilon(\theta)$ is functionally identical to \cref{eq:HTFIM_disp} and round brackets have been used to highlight the continuous nature of the argument. 
To find an explicitly $N$-independent form of $H(t<0)$, one can introduce rescaled operators: $\gamma(\theta)=\sqrt{N}\tilde{\gamma}(\theta)$. 
In these operators, the Hamiltonian takes the form 
\begin{equation}
    \lim_{N\to\infty}H(t<0)=\int_{-\pi}^\pi \frac{d\theta}{2\pi}\epsilon(\theta) \gamma^\dagger(\theta)\gamma(\theta),
\end{equation}
and the operators obey the anti-commutation relations
\begin{equation}
    \left\{\gamma(\theta),\gamma(\phi)\right\}=N\left\{\tilde{\gamma}(\theta),\tilde{\gamma}(\phi)\right\}=2\pi\delta(\theta-\phi).
\end{equation}
The corresponding real-space operators are 
\begin{equation}
    b_n=\frac{1}{N}\int_{-\pi}^\pi \frac{d\theta}{\Delta_\theta}e^{-i\theta n}\gamma(\theta)=\int_{-\pi}^\pi \frac{d\theta}{2\pi} e^{-i\theta n}\gamma(\theta),
\end{equation}
and obey the anti-commutation relations
\begin{equation}
    \left\{\tilde{b}_n,\tilde{b}^\dagger_m\right\}=\delta_{n,m}.
\end{equation}
Due to the finite lattice spacing, the inverse Fourier transformation remains discrete
\begin{equation}
        \gamma(\theta)=\sum_{n} e^{in\theta}b_n.
\end{equation}
In the thermodynamic limit, the post-quench Hamiltonian can also be written in the fermion basis \cite{Rutkevich1999}
\begin{equation}
    H(t>0)=H(t<0)+V
\end{equation}
with 
\begin{equation}
    V=J\abs{h_\parallel} M \sum_{n}:\exp\left(-\sum_{j<n}^N\psi^{(+)}_j\psi_j^{(-)}\right):,
\end{equation}
where $:\mathcal{O}:$ symbolizes normal-ordering and 
\begin{equation}
\begin{aligned}
    \psi^{(+)}_j&=i\int_{-\pi}^\pi\frac{d\theta}{\sqrt{ 2\pi}}\frac{e^{ij\theta}}{\sqrt{\epsilon(\theta)}}\left[\gamma(\theta)+\gamma^\dagger(-\theta)\right],\;
    \psi^{(-)}_j=i\int_{-\pi}^\pi\frac{d\theta}{\sqrt{ 2\pi}}e^{ij\theta}\sqrt{\epsilon(\theta)}\left[-\gamma(\theta)+\gamma^\dagger(-\theta)\right].
\end{aligned}
\end{equation}

One can split $V$ into two contributions: one that conserves the number of kinks $V_0$ and the rest which couples sectors with different number of kinks. 
The simplest way to achieve this is to define $V_0$ in the flat-dispersion limit \cite{Rutkevich1999}
\begin{equation}\label{eq:V0}
    V_0=\lim_{h_\perp\to0}V=J\abs{h_\parallel} M \sum_{n}:\exp\left(-2\sum_{j<n} b^\dagger_jb_j\right):.
\end{equation}
One can thus reorder the post-quench Hamiltonian into the form
\begin{equation}
    H(t>0)=H_0+\tilde{V},
\end{equation}
where $H_0=H(t<0)+V_0$ and $\tilde{V}=V-V_0$. 
As $V_0$ does not couple sectors with a different number of kinks, the coupling between sectors with different fermion number is fully described by $\tilde{V}$. 
One notices that since $\tilde{V}$ is a normal-ordered exponential of an even number of $\gamma$'s, it can only add or remove an even number of fermions. 
Considering quenches from $\ket{0_+}$ means that for $t<0$ there are no fermions in the system and for $t>0$ one will thus only find states with an even number of fermions. 

\subsection{Two-fermion states}\label{sec:two_fermion}
If one considers either a small longitudinal field or a small transverse field, then $\tilde{V}$ only weakly couples sectors with different fermion numbers.
This creates a time scale on which the system can be approximated by a truncated Hilbert space consisting of the ground state and a single two-fermion state. 
The two-fermion state has to conserve momentum and we can thus describe the $L$'th two-fermion state with zero total momentum as 
\begin{equation}
    \ket{\phi_L}=\mathcal{N}\int_{-\pi}^\pi d\theta\phi_L(\theta)\gamma^\dagger(\theta)\gamma^\dagger(-\theta)\ket{0_+}.
\end{equation}
These states are eigenstates of the two-body eigenvalue problem (EVP)
with energies $E_L$
\begin{equation}\label{eq:2FermionEVP}
    H_0 \ket{\phi_L}=E_L\ket{\phi_L}.
\end{equation}
The normalization $\mathcal{N}$ is defined by the overlap
\begin{equation}
    \braket{\phi_L}{\phi_{L'}}=\mathcal{N}^2\int_{-\pi}^\pi d\theta \,d\phi\,\phi_L^*(\theta)\phi_{L'}(\phi)\bra{0_+}\gamma(-\theta)\gamma(\theta)\gamma^\dagger(\phi)\gamma^\dagger(-\phi)\ket{0_+}=1.
\end{equation}
Using the symmetric momentum interval and the Pauli exclusion principle, it follows that $\phi_L(\theta)=-\phi_L(-\theta)$. 
Additionally, our numerical solution of the two-body problem is done for a finite $N$, which we ensure is large enough to guarantee converged results.
All together, this gives the normalization $\mathcal{N}=1/\sqrt{4\pi\Delta_\theta N^2}=1/\sqrt{2(2\pi)^2N}$, such that the momentum-space wave function takes the form
\begin{equation}
    \ket{\phi_L}=\frac{1}{\sqrt{2N}}\int_{-\pi}^\pi \frac{d\theta}{2\pi}\phi_L(\theta)\gamma^\dagger(\theta)\gamma^\dagger(-\theta)\ket{0_+}.
\end{equation}
The momentum-space representation of the two-fermion state can easily be transformed to real space
\begin{equation}
\begin{aligned}
    \ket{\phi_L}&=\frac{1}{\sqrt{2N}}\sum_{n,m,m'}\int_{-\pi}^\pi \frac{d\theta}{2\pi}\exp\left[i\theta\left(n+m-m'\right)\right]\phi_{L,n}b^\dagger_{m'}b^\dagger_m\ket{0_+}\\
    &=\frac{1}{\sqrt{2N}}\sum_{n,m}\phi_{L,n}b^\dagger_{m+n}b^\dagger_m\ket{0_+},\\
    \end{aligned}
\end{equation}
which explicitly highlights the translation-invariant (i.e. zero-total-momentum) nature of the two-fermion state. 

The two-fermion EVP in \cref{eq:2FermionEVP} takes a convenient form in the real-space representation, since $V_0$ in \cref{eq:V0} is diagonal \cite{Rutkevich1999,Wilczek2023}
\begin{equation}\label{eq:EVP}
    \sum_{m\in Z} K_{n-m} \phi_{L,m}-\alpha n \phi_{L,n}=E_L \phi_{L,n},
\end{equation}
with $\alpha=2J M\abs{h_\parallel}$ and the kinetic part given by the Fourier transform of the dispersion
\begin{equation}\label{eq:discrete_kinetic_kernel}
    K_{n}=\int_{-\pi}^{\pi}\frac{d\theta}{\pi}\epsilon(\theta)e^{i n\theta}.
\end{equation}

\subsubsection{Coupling to FV and magnetization}
Due to $\tilde{V}$ the two-fermion states are coupled to the FV.
One can compute this coupling analytically \cite{Rutkevich1999,Wilczek2023}
\begin{equation}\label{eq:FVbub_overlap}
    \Omega_L\equiv\bra{0_+}V\ket{\phi_L}=\frac{1}{2}\sqrt{N}M J\abs{h_\parallel}\sum_{n=1}^N h_\perp^n\phi_{L,n}=\Omega_L^*.
\end{equation}
By definition the average longitudinal magnetization can be written as
\begin{equation}
  \frac{1}{N}\left<\sum_{n=1}^N\sigma^x_n\right>=\frac{\left<V\right>}{NJ \abs{h_\parallel}}.
\end{equation}
Within the truncated Hilbert space spanned by the FV state and the 
one-bubble states, one can approximate the average magnetization operator as 
\begin{equation}\label{eq:singleBody_magOP_0}
    m_{avg}= M\dyad{0_+} +\frac{1}{NJ \abs{h_\parallel}}\sum_{L=1}^N\Omega_L\left(\dyad{0_+}{\phi_L}+\dyad{\phi_L}{0_+}\right)+\frac{1}{N J\abs{h_\parallel}}\sum_{L,L'=1}^N\tilde{\lambda}_{L,L'}\dyad{\phi_L}{\phi_{L'}},
\end{equation}
where we have defined
\begin{equation}
\bra{\phi_L}V\ket{\phi_{L'}}\approx\bra{\phi_L}V_0\ket{\phi_{L'}}=\tilde{\lambda}_{L,L'}.
\end{equation}

The overlap $\tilde{\lambda}_{L,L'}$ is straightforward to compute as $V_0$ is diagonal in the real-space representation and takes the form
\begin{equation}\label{eq:lambda_tilde}
\begin{aligned}
    \tilde{\lambda}_{L,L'}=&J\abs{h_\parallel} MN\sum_{n=1}^N\phi_{L,n}^*\phi_{L',n}\left(1-\frac{2n}{N}\right)\\=&J\abs{h_\parallel} MN\delta_{L,L'}-J\abs{h_\parallel} MN\sum_{n=1}^N\phi_{L,n}\phi_{L',n}\frac{2n}{N},
\end{aligned}
\end{equation}
where we used the fact that the spatial coefficients are real, making $\tilde{\lambda}_{L,L'}$ real and symmetric. 
Upon inserting $\tilde{\lambda}_{L,L'}$ into the magnetization in \cref{eq:singleBody_magOP_0} one notices that the constant terms can be combined into an identity operator such that $m_{avg}$ can be written as
\begin{equation}\label{eq:singleBody_magOP}
    m_{avg}= \mathds{1}\, M+\frac{1}{N J\abs{h_\parallel}}\sum_{L=1}^N\Omega_L\left(\dyad{0_+}{\phi_L}+\dyad{\phi_L}{0_+}\right)+\frac{1}{NJ \abs{h_\parallel}}\sum_{L,L'=1}^N\lambda_{L,L'}\dyad{\phi_L}{\phi_{L'}},
\end{equation}
with 
\begin{equation}\label{eq:lambda}
\lambda_{L,L'}=-J\abs{h_\parallel} 2M\sum_{n=1}^Nn\phi_{L,n}\phi_{L',n}.
\end{equation}

\subsection{Many-body theory for translation-invariant bubbles}\label{sec:CBT}
As the system is translation-invariant and infinitely large, a single bubble does not affect the magnetization of the system. 
This is clearly seen by $\lim_{N\to\infty}\lambda_{L,L'}/N=0$ and
$\lim_{N\to\infty}\Omega_{L}/N=0$. 
This means that a theory with only a single bubble can only be a good description of a large system when the fluctuations on top of the FV are extremely small. This approximation was done, e.g., in \cite{Wilczek2023}. 
Otherwise, it is necessary to use a many-body description, which allows for a finite bubble-density in the thermodynamic limit. 

The simplest conceivable many-body theory is based on the infinite size of the system and on the fact that the one-bubble states $\ket{\phi_L}$ consists of superpositions of bubbles with finite sizes. 
For low enough densities one thus expects that the state can be approximated as multiple non-interacting bubbles.
Each bubble consists of two fermions, so as long as the bubbles do not interact with each other it should be valid to think of the bubbles as non-interacting bosons.
This idea belongs to the broader scope of methods known as cobosonization \cite{Tichy2014} and is related to the fermionic squeezed-state approximation introduced in \cite{Scopa:PRB2022}.  

Within this method, we define a creation operator for the $L$'th bubble state
\begin{equation}
    d^\dagger_L=\frac{1}{\sqrt{2N}}\int_{-\pi}^\pi\frac{d\theta}{2\pi}\phi_L(\theta)\gamma^\dagger(\theta)\gamma^\dagger(-\theta),
\end{equation}
The fermionic anti-commutation relations in \cref{eq:fermionAC_finite} lead  to
\begin{equation}
\left[d^\dagger_L,d^\dagger_{L'}\right]=\Big[d_L,d_{L'}\Big]=0,
\end{equation}
while
\begin{equation}
\begin{aligned}\label{eq:TI_Commutation}
    \left[d_L,d^\dagger_{L'}\right]=&\frac{1}{2N}\int_{-\pi}^\pi\frac{d\theta}{(2\pi)^2}\Big[(N\Delta_\theta)^2\delta(0)\phi_L^*(\theta)\phi_{L'}(\theta)-(N\Delta_\theta)^2\delta(0)\phi_L^*(\theta)\phi_{L'}(-\theta)-N\Delta_\theta\phi_L^*(\theta)\phi_{L'}(\theta)\gamma^\dagger(-\theta)\gamma(-\theta)\\&\hspace{10mm}+N\Delta_\theta\phi_L^*(\theta)\phi_{L'}(-\theta)\gamma^\dagger(-\theta)\gamma(-\theta)-N\Delta_\theta\phi_L^*(\theta)\phi_{L'}(\theta)\gamma^\dagger(\theta)\gamma(\theta)+N\Delta_\theta\phi_L^*(\theta)\phi_{L'}(-\theta)\gamma^\dagger(\theta)\gamma(\theta)\Big]\\
    =&\delta_{L,L'}-\frac{2}{N}\int_{-\pi}^\pi \frac{d\theta}{2\pi}\phi_L^*(\theta)\phi_{L'}(\theta)\gamma^\dagger(\theta)\gamma(\theta),
\end{aligned}
\end{equation}
where the anti-symmetry of $\phi_L(\theta)$ has been exploited. 
This result clearly shows that bosonic commutation relations emerge when only one bubble is present in the system. 

In the real-space representation the commutation relations take the form
\begin{equation}
    \left[d_L,d^\dagger_{L'}\right]=\delta_{L,L'}-\frac{2}{N}\sum_{n,m,m'=1}^N\phi_{L,n}^*\phi_{L',m}b^\dagger_{m-n+m'}b_{m'}.
\end{equation}
Exploiting the translation-invariance by doing the sum over $m'$ one finds
\begin{equation}\begin{aligned}
\left[d_L,d^\dagger_{L'}\right]&=\delta_{L,L'}-2\sum_{n,m}^N\phi_{L,n}^*\phi_{L',m}b^\dagger_{m-n}b_0.
\end{aligned}
\end{equation}

In the $h_\perp\to0$ case the fermion wave functions become strongly localized $\phi_{L,n}\approx\delta_{L,n}$.
In this case, the perturbation to the bosonic commutator becomes 
\begin{equation}
\lim_{h_\perp\to0}\left[d_L,d^\dagger_{L'}\right]=\delta_{L,L'}-2 b^\dagger_{L-L'}b_0,
\end{equation}
which is always unity for $L=L'$ but is related to the fermion two-point correlation function for $L\neq L'$. 
\subsubsection{Coherent bubble theory}
We will now assume that the density of bubbles is small enough that the perturbation of the bosonic commutation relations in \cref{eq:TI_Commutation} can be neglected. 
Under this assumption, the state with $n$ bubbles of type $L$ and $m$ bubbles of type $L'$ is given as 
\begin{equation}
    \ket{n_L,m_{L'}}=\frac{1}{\sqrt{n!m!}}\left(d_L^\dagger\right)^n\left(d_{L'}^\dagger\right)^m\ket{0_+}.
\end{equation}
The Hamiltonian contains the single-bubble energies and their coupling to the FV
\begin{equation}
H=\sum_{L=1}^NE_Ld^\dagger_Ld_L+\Omega_L\left(d_L+d^\dagger_L\right).
\end{equation}
The non-interacting nature of the bosons means that we can diagonalize the Hamiltonian using a multi-mode displacement operator
\begin{equation}
    D = \bigotimes_LD_L(\beta_L),
\end{equation}
with the single-mode displacement operator being defined as 
\begin{equation}
    D_L(\beta)=\exp\left(\beta d_L^\dagger-\beta^* d_L\right).
\end{equation}
This operator is unitary with 
\begin{equation}
D_L^\dagger(\beta)=D_L^{-1}(\beta)=D_L(-\beta),    
\end{equation}
and its action on the annihilation and creation operators are simple shifts
\begin{equation}
\begin{aligned}
    D_L^\dagger(\beta)d_LD_L(\beta)&=d_L+\beta\\
    D_L^\dagger(\beta)d_L^\dagger D_L(\beta)&=d_L^\dagger+\beta^*.
\end{aligned}
\end{equation}
When the displacement operator is applied to the vacuum state it generates a coherent state
\begin{equation}
    D_L(\beta_L)\ket{0_+}=\exp\left(-\frac{1}{2}\abs{\beta_L}^2\right)\sum_{n=0}^N\frac{\beta_L^n}{\sqrt{n!}}\ket{n_L}=\ket{\beta_L}.
\end{equation}
As the coherent states constitute an over-complete basis, the overlap of two coherent states is finite
\begin{equation}\label{eq:overlap_coh}
    \abs{\ip{\beta_L}{\gamma_L}}^2=\exp\left(-\abs{\beta_L-\gamma_L}^2\right).
\end{equation}
Applying the multi-mode displacement operator to the Hamiltonian, one finds 
\begin{equation}
    \begin{aligned}
        \tilde{H}&=D^\dagger HD\\
        &=\sum_L E_L\left(d^\dagger_L+\beta_L^*\right)\Big(d_L+\beta_L\Big)+\Omega_L\left(d^\dagger_L+d_L+2\text{Re}(\beta_L)\right)\\
        &=\sum_L E_L d^\dagger_Ld_L+E_L\abs{\beta_L}^2+2\text{Re}(\beta_L)\Omega_L+\left(\Omega_L+E_L\beta_L^*\right)d_L+\left(\Omega_L+E_L\beta_L\right)d_L^\dagger,
    \end{aligned}
\end{equation}
where we will use the tilde to indicate that we are in the displaced basis. 
The linear terms can be removed by choosing $\beta_L=-\Omega_L/E_L$, which makes the Hamiltonian diagonal
\begin{equation}
    \tilde{H}=\sum_L E_L d^\dagger_L d_L-\frac{\Omega_L^2}{E_L}.
\end{equation}
The displaced Hamiltonian is a simple harmonic oscillator where each mode acquires a shift. 
As the modes do not interact, this shift only leads to an overall phase of the wave-function $\sum_L\Omega_L^2/E_L=v$. 
Any time evolved state can therefore be represented as 
\begin{equation}
    \ket{\tilde{\psi}(t)}=e^{it v}\sum_{\{n_L\}}c_{\{n_L\}}\exp\left(-i\sum_LE_Ln_L t\right)\ket{n_1,n_2,...,n_N}=e^{it v}\sum_{\{n_L\}}c_{\{n_L\}}\bigotimes_L\exp\left(-iE_Ln_L t\right)\ket{n_L}.
    \end{equation}
For our purpose, we are interested in time evolving a specific state, namely, the false vacuum state. 
Transforming the FV to the diagonal basis is straightforward
\begin{equation}
   \begin{aligned}
       \ket{\tilde{\psi}(0)}=D^\dagger\ket{0_+}=\bigotimes_LD_L(\Omega_L/E_L)\ket{0_+}=\bigotimes_L\ket{\Omega_L/E_L}=\bigotimes_L\exp\left[-\frac{1}{2}\left(\frac{\Omega_L}{E_L}\right)^2\right]\sum_{n=0}^N\frac{\left(\frac{\Omega_L}{E_L}\right)^n}{\sqrt{n!}}\ket{n_L},
   \end{aligned} 
\end{equation}
The coefficients in the wave-function expansion now take the form 
\begin{equation}
    \ip{\tilde{\psi}(0)}{n1,n2,...,n_N}=\prod_L\exp\left[-\frac{1}{2}\left(\frac{\Omega_L}{E_L}\right)^2\right]\frac{\left(\frac{\Omega_L}{E_L}\right)^{n_L}}{\sqrt{n_L!}}.
\end{equation}
As these coefficients are just a product over the modes, the entire time-evolved state is a product state for all times
\begin{equation}\label{eq:timeDep_state_app}
\begin{aligned}
    \ket{\tilde{\psi}(t)}&=e^{itv}\bigotimes_L\exp\left[-\frac{1}{2}\left(\frac{\Omega_L}{E_L}\right)^2\right]\sum_{\{n_L\}}\frac{\left(\frac{\Omega_L}{E_L}e^{-iE_L t}\right)^{n_L}}{\sqrt{n_L!}}\ket{n_L}\\&=e^{itv}\bigotimes_L\ket{\frac{\Omega_Le^{-iE_L t}}{E_L}}.
    \end{aligned}
\end{equation} 
Due to the unitarity of the displacement operator, the state is trivially normalized for all times. 
Displacing this state to the original basis, one immediately finds \cite{Glauber1963}
\begin{equation}\label{eq:chiAmp}
   \ket{\psi(t)}= D\ket{\tilde{\psi}(t)}=e^{itv}\bigotimes_L\ket{\frac{\Omega_L}{E_L}\left(e^{-iE_L t}-1\right)},
\end{equation}
from which we identify $\chi_L=\Omega_L\left(e^{-iE_L t}-1\right)/E_L$, introduced in the main text.
\subsubsection{Dynamics of excitations}
The number of excitations in the $L$'th state is given by 
\begin{equation}
\begin{aligned}
    \bra{\psi(t)}d_L^\dagger d_L\ket{\psi(t)}&=\bra{\tilde{\psi}(t)}D^\dagger d_L^\dagger D D^\dagger d_LD\ket{\tilde{\psi}(t)}\\&=\bra{\tilde{\psi}(t)} \left(d_L^\dagger-\frac{\Omega_L}{E_L}\right) \left(d_L-\frac{\Omega_L}{E_L}\right)\ket{\tilde{\psi}(t)}\\
    &=\left(\frac{2\Omega_L}{E_L}\right)^2\sin^2\left(E_Lt/2\right),
    \end{aligned}
\end{equation}
which means that the system initially has no occupation and is therefore consistent with the initial state used. 
With this result our prediction is that the dynamics of the $L$'th excitation has a time-scale directly set by $1/E_L$ but that its weight in the dynamics is set by $\left(\frac{\Omega_L}{E_L}\right)^2$. 
As the system is infinite, the relevant quantity is not the number of bubbles but instead their density 
\begin{equation}\label{eq:NL}
    \mathcal{N}_{L}=\frac{\bra{\psi(t)}d_L^\dagger d_L\ket{\psi(t)}}{N}=\frac{1}{N}\left(\frac{2\Omega_L}{E_L}\right)^2\sin^2\left(E_Lt/2\right).
\end{equation}
Since $\Omega_L\propto \sqrt{N}$ as seen in \cref{eq:FVbub_overlap} and $E_L$ is an intensive quantity, the density of the $L$'th excitation becomes intensive and therefore well-defined in the thermodynamic limit. 
This observable describes the average number of excitations of species $L$ at a single site. 
For the case of a resonant excitation, $E_r\to 0$ allowing us to expand the sine
\begin{equation}\label{eq:N_t2_growth}
    \mathcal{N}_{r}=\frac{\Omega_r^2 t^2}{N},
\end{equation}
giving the resonant physics a time scale of 
\begin{equation}
    T_{r}=\frac{2}{ JM \abs{h_\parallel}\sum_{n=1}^N h_\perp^n\phi_{r,n}}.
\end{equation}
Finally, we can consider the total density of excitations 
\begin{equation}\label{eq:totDen}
    \mathcal{N}=\sum_L \mathcal{N}_L=\frac{1}{N}\sum_L\left(\frac{2\Omega_L}{E_L}\right)^2\sin^2\left(E_Lt/2\right).
\end{equation}

\subsubsection{Average magnetization}\label{app:avg_mag}
To compare the results of the CBT with MPS calculations, we wish to compute an approximation to the average magnetization within the CBT.
To compute the magnetization, we can directly promote the single-body average magnetization operator in \cref{eq:singleBody_magOP} to a many body-operator
\begin{equation}
    m_{avg} = M +\frac{1}{NJ \abs{h_\parallel}}\sum_{L=1}^N\Omega_L\left(d_L+d^\dagger_L\right)+\frac{1}{N J\abs{h_\parallel}}\sum_{L,L'=1}^N\lambda_{L,L'}d_L^\dagger d_{L'},
\end{equation} 
where we remind the reader that there is an additional flat-band approximation in $\lambda_{L,L'}$.
Using the same procedure as for the bubble occupation, we transform $m_{avg}$ to the displaced basis
\begin{equation}
\begin{aligned}
    \tilde{m}_{avg}&=D^\dagger m_{avg}D,\\
    &=M+\frac{1}{NJ\abs{h_\parallel}}\sum_L\Omega_L
    \left(d_L+d^\dagger_L-2\frac{\Omega_L}{E_L}\right)+\frac{1}{NJ\abs{h_\parallel}}\sum_{L,L'}\lambda_{L,L'}
    \Bigg[d^\dagger_L d_{L'}-\frac{\Omega_L}{E_L}d_{L'}\\&\hspace{95mm}-\frac{\Omega_{L'}}{E_{L'}}d^\dagger_L+\frac{\Omega_L\Omega_{L'}}{E_L E_{L'}}\Bigg]\\
    &=M+G_\Omega+G_\lambda.
\end{aligned}
\end{equation}
We can now compute overlaps with the time-dependent state in \cref{eq:timeDep_state_app} of the two different terms. 
The first contribution arises from the coupling between the false vacuum and one-bubble states
\begin{equation}\label{eq:mag_crossTerms}
\begin{aligned}
    \langle G_\Omega\rangle(t)&=\bra{\psi(t)}G_\Omega\ket{\psi(t)}\\
    &=\frac{1}{N J\abs{h_\parallel}}\sum_L\Omega_L\left[\frac{\Omega_L e^{-it E_L}}{E_L}+\frac{\Omega_L e^{it E_L}}{E_L}-2\frac{\Omega_L}{E_L}\right]\\
    &=\frac{2}{N J\abs{h_\parallel}}\sum_L\frac{\Omega_L^2}{E_L}\left[\cos\left(E_L t\right)-1\right].\\
\end{aligned}
\end{equation}
This term is seen to be intensive and consistent with our initial state as the entire contribution vanishes at $t=0$. 

The second term is related to the magnetization of a bubble state 
\begin{equation}\label{eq:T_lambda}
\begin{aligned}
    \langle G_\lambda\rangle(t)&=\bra{\psi(t)}G_\lambda\ket{\psi(t)},\\
    &=\frac{1}{N J\abs{h_\parallel}}\sum_{L,L'}\lambda_{L,L'}\frac{\Omega_L\Omega_{L'}}{E_L E_{L'}}\Big(e^{i(E_L-E_{L'})t}-e^{-iE_{L'}t}-e^{iE_L t}+1\Big)\\
    &=\frac{1}{N J\abs{h_\parallel}}\sum_{L,L'}\lambda_{L,L'}\frac{\Omega_L\Omega_{L'}}{E_L E_{L'}}\Big(\cos\left\{(E_L-E_{L'})t\right\}-2\cos\left(E_{L}t\right)+1\Big),\\
\end{aligned}
\end{equation}
where we used the symmetric property of $\lambda_{L,L'}$ seen in \cref{eq:lambda}. 
As for $\langle G_\Omega\rangle$ also $\langle G_\lambda\rangle$ satisfies the consistency check that it vanishes at $t=0$. 
One notices that the extensive property of $\Omega_L$ is canceled and since $\lambda_{L,L'}$, in \cref{eq:lambda}, is intensive the average magnetization is indeed an intensive quantity. 
If the eigenstates are sufficiently far apart in energy, one expects that the physics is dominated by the resonant bubble after some initial dynamics. Assuming that an almost resonant bubble exists in the system such that we can expand the cosines, the slow envelope of the magnetization grows quadratically
\begin{equation}
    \lim_{E_r\to0}\langle m_{avg}\rangle(t)=M-\frac{\Omega_r^2 t^2}{N J\abs{h_\parallel}}\left(E_R+\lambda_r\right)=M-\frac{\lambda_r\Omega_r^2 t^2}{N J\abs{h_\parallel}}.
\end{equation}

\subsubsection{Dynamics of bubbles}
The density of bubbles of size $n$ is straightforward to compute within the single-bubble theory 
\begin{equation}\
    P_n = \frac{1}{N}\abs{\bra{0_+}\sum_m b_mb_{n+m}\ket{\psi(t)}}^2
\end{equation}
as the operators $b_n$ action on $\ket{\psi}$ is well-understood.
In the many-bubble theory the wave-function has the form 
\begin{equation}
    \ket{\psi(t)}=\sum_{m,L} c_{m,L}\left(d_L^\dagger\right)^m\ket{0_+}.
\end{equation}
The action of $b_n$ on the coherent bubble states is highly non-trivial and we can therefore not directly compute $P_n$.
However, if we neglect the quantum cross-terms then we can write an approximation for the overlap between 
states with fermions a distance $n$ apart and our coherent states given by
\begin{equation}
    \braket{n}{\psi} = \frac{1}{\sqrt{N}}\sum_L \langle d_L^\dagger\rangle \phi_L(n).
\end{equation}
Within this approximation $P_n$ is given by
\begin{equation}\label{eq:Pn}
\begin{aligned}
    P_n&=\frac{1}{N}\sum_{L,L'} \langle d_L^\dagger\rangle \langle d_{L'}\rangle \phi_L(n)\phi_{L'}^*(n)\\
    &= \sum_{L,L'}\Gamma_{L,L'}(n)\Big\{\cos\left[t(E_L-E_{L'}|)\right]-2\cos(t E_L)+1\Big\},
\end{aligned}
\end{equation}
where
\begin{equation}
    \Gamma_{L,L'}(n)=\frac{1}{N}\phi_L(n)\phi_{L'}^*(n)\frac{\Omega_L \Omega_{L'}}{E_L E_{L'}}.
\end{equation}
As a sanity check we can compute the total density of bubbles in the system
\begin{equation}
    \begin{aligned}
        \sum_nP_n&=\frac{1}{N}\sum_{L,L',n} \langle d_L^\dagger\rangle \langle d_{L'}\rangle \phi_L(n)\phi_{L'}^*(n)\\
        &=\frac{1}{N}\sum_{L,L'} \langle d_L^\dagger\rangle \langle d_{L'}\rangle \delta_{L,L'}\\
        &=\frac{1}{N}\sum_L \langle d_L^\dagger d_L\rangle\\
        &=\mathcal{N},
    \end{aligned}
\end{equation}
 with the excitation density defined in \cref{eq:totDen}. 
 So the number-density of spatial bubbles is equivalent to the total number-density of bosonic excitations as one would expect from our simple non-interacting theory. 

To show that some care has to be taken with our approximation of $P_n$ we can use it to compute the magnetization. 
Within our approximate picture and with only a small excitation density, the magnetization should be related to the density of TV domains through
\begin{equation}
    m(t)\approx M\left[1-2\rho(t)\right]=M\left[1-2\sum_n n P_n(t)\right].
\end{equation}
Writing out the last term
\begin{equation}
\begin{aligned}
    -2M \sum_n n P_n(t)&=-2M \sum_{n,L,L'} n\Gamma_{L,L'}(n) F(t)\\
    &=-2M \sum_{n,L,L'} \frac{n\phi_L(n)\phi_{L'}^*(n)}{N}\frac{\Omega_L \Omega_{L'}}{E_L E_{L'}} F(t),\\
\end{aligned} 
\end{equation}
 where $F(t)$ is the time-dependent function in the curly brackets in \cref{eq:Pn}.
 Writing out the $\lambda$-contribution to the magnetization in \cref{eq:T_lambda} one finds
 \begin{equation}
 \begin{aligned}
     \langle G_\lambda\rangle(t)=\sum_{L,L'}\frac{-2 M n \phi_L(n)\phi_{L'}^*(n)}{N}\frac{\Omega_L\Omega_{L'}}{E_L E_{L'}}F(t)=-2M \sum_n n P_n(t).
 \end{aligned}
 \end{equation}
 This means that the error made in using $P_n$ to compute the magnetization is due to the fact that the quantum cross-terms from $G_\Omega$ in \cref{eq:mag_crossTerms} are missing. 
So quantities computed with $P_n$ generically include additional approximations on top of the ones already present in CBT. 
\begin{figure}
    \centering
\includegraphics[width=.5\linewidth]{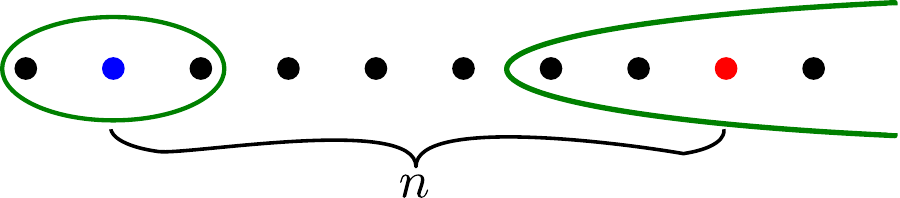}
    \caption{Sketch illustrating how to compute the correlation function. The green ovals represent TV bubbles.}
    \label{fig:corr}
\end{figure}
\subsubsection{spin-spin correlation function}
To understand how the spatial structure in the fermion states translate into spatial structure in the spin chain we seek to approximate the spin-spin correlation function given by
\begin{equation}
    C_{n}(t)=\frac{1}{N}\sum_{j=1}^N\left(\langle\sigma^x_{n+j}\sigma^x_j\rangle-\langle\sigma^x_j\rangle^2\right).
\end{equation}
To approximate $C_n(t)$ we first compute the probability, $Q_n$, of two spins having opposite magnetization when being separated by $n$ sites which is sketched in \cref{fig:corr} as the probability that the red and blue site have opposite magnetization. 
Consistent with our assumption that the excitation density is small we consider the probability in the case when only one bubble is present within this region.
There are two different scenarios that must be accounted for. 
The first is illustrated around the red site in \cref{fig:corr} (ignore bubble around blue site). 
In this scenario the bubble size $l_B$ is large than the separation between the sites: $l_B>n$. 
In this case there are $n$ different locations for the bubble such that the red site is covered but the blue site is not. 
The opposite regime of $l_B<n$ is illustrated around the blue site (ignore the previously discussed bubble around the red site). 
In this case there are $l_B$ different locations where the blue site is covered and thus gives rise to the opposite magnetization of the two colored sites. 
Putting this together, we can write the probability of the two spins having opposite magnetization as
\begin{equation}
    Q_n(t)=2\sum_l \textrm{min}(n,l) P_l(t),
\end{equation}
where the factor of two arises from the fact that we also have to include the processes where the two sites are swapped (large bubble from the left and small bubble to right). 

With this object, we are able to compute the total spin-spin correlation function as a sum of two contributions.
The first is the probability of the two spins pointing in the same direction, and the second is the probability of them being opposite
\begin{equation}
\begin{aligned}
    \mathcal{C}_n(t)&=\frac{1}{N}\sum_{j=1}^N\langle\sigma^x_{n+j}\sigma^x_j\rangle(t)\\&=M^2\left[1-Q_n(t)\right]+M^2\left[-Q_n(t)\right]\\
    &=M^2\left[1-2Q_n(t)\right]\\
    &=M^2\left[1-4\sum_l \textrm{min}(n,l) P_l(t)\right].
\end{aligned}
\end{equation}
To compute the disconnected contribution, we consider the limit of $\lim_{n\to\infty}$
\begin{equation}
    \begin{aligned}
        \lim_{n\to\infty}\mathcal{C}_n(t)&=\frac{1}{N}\sum_{j}\langle\sigma^x_n\rangle\langle\sigma^x_j\rangle\\
        &=\frac{1}{N}\sum_{j}\langle\sigma^x_j\rangle^2\\
        &=M^2\left[1-4\sum_l l P_l(t)\right]\\
        &=M^2\left[1-4\rho(t)\right],
    \end{aligned}
\end{equation}
where we have exploited the translation invariance of the system.
With these results, we can approximate the connected spin-spin correlation function as
\begin{equation}\label{eq:CorrCBT}
    C_n(t)=4M^2\sum_{l=n+1}^N (l-n) P_l(t).
\end{equation}
The approximation has the simple intuitive structure that only bubbles larger than the distance between the sites contribute to the connected correlations. 

Lastly we remark that for both the magnetization and the correlation function we have neglected the delocalized nature of the domain walls, which is only valid if the relevant fermion bubbles are much larger than the domain wall spreading in $W$ discussed in the main text.
For the considered parameters in the main text this is the case as for the rigid parameters $W_\rho\approx0.26$ where the resonant bubble is of size $4$, while for the expandable regime $W_\chi\approx3.46$ where the bubbles extend up to more than a hundred lattice sites (see plot of $\abs{\phi_{r}(n)}^2$ in main text).

\subsection{Scaling limit}\label{sec:scaling}
So far we have considered a theory with lattice spacing $a=1$. In the main text we discussed how the finite spacing leads to features in the dynamics such as the Bloch oscillations. In this section we will discuss how decreasing $a$ while keeping $X=a N$ fixed and correspondingly scaling the external fields makes it possible to reach the continuum limit. Throughout the section we choose $\hbar=1$ and $c=1$. The first part of this section has a significant overlap with the work done in \cite{Wilczek2023}. 

The starting point for this scaling is requiring that the low-energy behavior of the fermion dispersion \cref{eq:HTFIM_disp} is invariant as $a$ is decreased. 
The lattice spacing emerges in the theory when introducing the physical momentum $k$: $\theta=k a$. 
While the quasi-momentum $\theta$ is unaffected by the scaling, the Brillouin zone related to the physical momentum $k$ grows inversely with $a$ such that $k\in\{-\pi/a,\pi/a\}$.
Writing the dispersion relation in terms of the physical momentum and expanding for small $k$
\begin{equation}\label{eq:expand_disp}
    \begin{aligned}
        \epsilon(k a)&=2J \sqrt{1 - 2 h_\perp  \cos\left(k a\right) + h_\perp^2}\\
        &\approx 2J\sqrt{(1-h_\perp)^2+h_\perp a^2k^2}\\
        &\approx 2J(1-h_\perp)+\frac{J h_\perp a^2k^2}{1-h_\perp}.
    \end{aligned}
\end{equation}
The first term is the energy gap $\epsilon(0)=\Delta$ that must be overcome to create a kink. 
To keep $\Delta$ invariant while varying the energy scale $J$ the transverse field has to be tuned according to
\begin{equation}\label{eq:hperp_scaling1}
    h_\perp=1-\frac{\Delta}{2J}.
\end{equation}
Besides fixing the gap we also need to fix the mass of the excitations, which can be extracted from the second term of \cref{eq:expand_disp}
\begin{equation}\label{eq:mass_scaling}
\begin{aligned}
    \frac{1}{2m}&=\frac{J h_\perp a^2}{1-h_\perp}\\
    &=\frac{2 J^2 h_\perp a^2}{\Delta}\\
    &=\frac{2 J^2 a^2}{\Delta}-Ja^2,\\
\end{aligned}
\end{equation}
where we have used \cref{eq:hperp_scaling1} to eliminate $h_\perp$.
To keep the mass scale-invariant, $J$ must be proportional to $a^{-1}$, which results in the second term of \cref{eq:mass_scaling} vanishing in the scaling limit. The numerical proportionality constant determines the mass of the fermions which we, for simplicity, choose to be $m=\Delta$ realized by the choice $J=1/2a$. 
This means that our theory scales to the Ising field theory in the continuum limit and the low-energy part of the dispersion is of the non-relativistic form 
\begin{equation}
    \epsilon(k a)\approx\Delta +\frac{k^2}{2\Delta},
\end{equation}
with the transverse field scaling as 
\begin{equation}\label{eq:hperp_scaling}
    h_\perp = 1- a\Delta.
\end{equation}

Within the non-relativistic approximation for the dispersion, one can scale the single-bubble EVP in \cref{eq:EVP}. 
To do this, we introduce the spatial variable $x=n a$ which leads to the two body problem
\begin{equation}\label{eq:s1_EVp}
    \sum_{y} K(x-y) \phi_{L}(y/a)-\frac{\alpha}{a} x \phi_{L}(x/a)=E_L \phi_{L}(x/a),
\end{equation}
where the kinetic kernel follows directly from \cref{eq:discrete_kinetic_kernel} and takes the  
\begin{equation}\begin{aligned}
    K(x-y)&=2 a\int_{-\pi/a}^{\pi/a} \frac{d k}{2\pi}\epsilon(k a)e^{i k (x-y)}\\
    &\approx 2 \Delta a \delta(x-y)+\frac{a}{\Delta}\int_{-\pi/a}^{\pi/a} \frac{d k}{2\pi}k^2e^{i k (x-y)},
\end{aligned}
\end{equation}
where we are considering the limit $a\to0$.
Inserting this kernel into the first term of \cref{eq:s1_EVp} one finds
\begin{equation}
    \begin{aligned}
        \sum_{y} K(x-y) \phi_{L}(y/a)&\approx 2 \Delta \phi_L(x/a)+\frac{1}{\Delta}\int_{-X/2}^{X/2}dy \int_{-\pi/a}^{\pi/a}\frac{dk}{2\pi} k^2 e^{ik (x-y)}\phi_L(y/a)\\
        &\approx 2 \Delta \phi_L(x/a)-\frac{1}{\Delta}\partial_x^2\phi_L(x/a).
    \end{aligned}
\end{equation}
Combining this with the two other terms in the single-bubble EVP in \cref{eq:s1_EVp} one finds its form in the $a\to0$ limit as
\begin{equation}\label{eq:s2_EVP}
    2 \Delta \psi_L(x)-\frac{1}{\Delta}\partial_x^2\psi_L(x)-\frac{\alpha}{a} x \psi_{L}(x)=E_L \psi_{L}(x).
\end{equation}
In this expression, we have identified the continuum limit of the wave function
\begin{equation}\label{eq:wf_scaling}
   \lim_{a\to0}\frac{\phi_L(x/a)}{\sqrt{a}}=\psi_L(x), 
\end{equation}
which is normalized according to
\begin{equation}
    \int_{0}^{X} dx \,\psi_L(x)\psi_{L'}(x)=\delta_{L,L'},
\end{equation}
where the odd-symmetry of $\psi_L(x)$ has been exploited to change the integral limits to match the discrete case discussed in the previous sections. 
The only term that explicitly depends on $a$ is the repulsive interaction. 
To ensure that this term has a finite value in the scaling limit, we need to require that $\alpha/a=\delta$ is independent of $a$. 
Using the definition of $\alpha$ below \cref{eq:EVP} we find an expression for $\delta$ of the form
\begin{equation}
    \delta=\frac{\alpha}{a}=\frac{2 J M \abs{h_\parallel}}{a}=\frac{M \abs{h_\parallel}}{a^2}.
\end{equation}
The average magnetization per site for the FV can be written in terms of $a$ and $\delta$ using the scaling relations for $h_\perp$
\begin{equation}\begin{aligned}
    M=&\left(1-h_\perp^2\right)^{1/8}\\=&\left[a \Delta\left(1+h_\perp\right)\right]^{1/8}\\=&\left(2 a \Delta-\left(a\Delta\right)^2\right)^{1/8}\\
    \approx& \left(2a\Delta\right)^{1/8}.
\end{aligned}
\end{equation}
Inserting this into $\delta$ one finds an expression that only contains $a,\,\Delta$ and $h_\parallel$
\begin{equation}
    \begin{aligned}
        \delta&=\frac{M \abs{h_\parallel}}{a^2}\\
        &=\frac{\left[a \Delta\left(2-a\Delta\right)\right]^{1/8} \abs{h_\parallel}}{a^2}\\
        &= \left[a^{-15} \Delta\left(2-a\Delta\right)\right]^{1/8} \abs{h_\parallel}\\
        &\approx (2\Delta)^{1/8} a^{-15/8}\abs{h_\parallel}.
    \end{aligned}
\end{equation}
Requiring $\delta$ to be $a$-independent gives the scaling of the longitudinal field 
\begin{equation}\label{eq:hlon_scaling}
\begin{aligned}
    \abs{h_\parallel}&=\delta\left[a^{-15} \Delta\left(2-a\Delta\right)\right]^{-1/8}\\
    &\approx \frac{\delta \ a^{15/8}}{(2\Delta)^{1/8}}.
\end{aligned}
\end{equation}
\begin{figure*}
    \centering
    \includegraphics[width=0.8\linewidth]{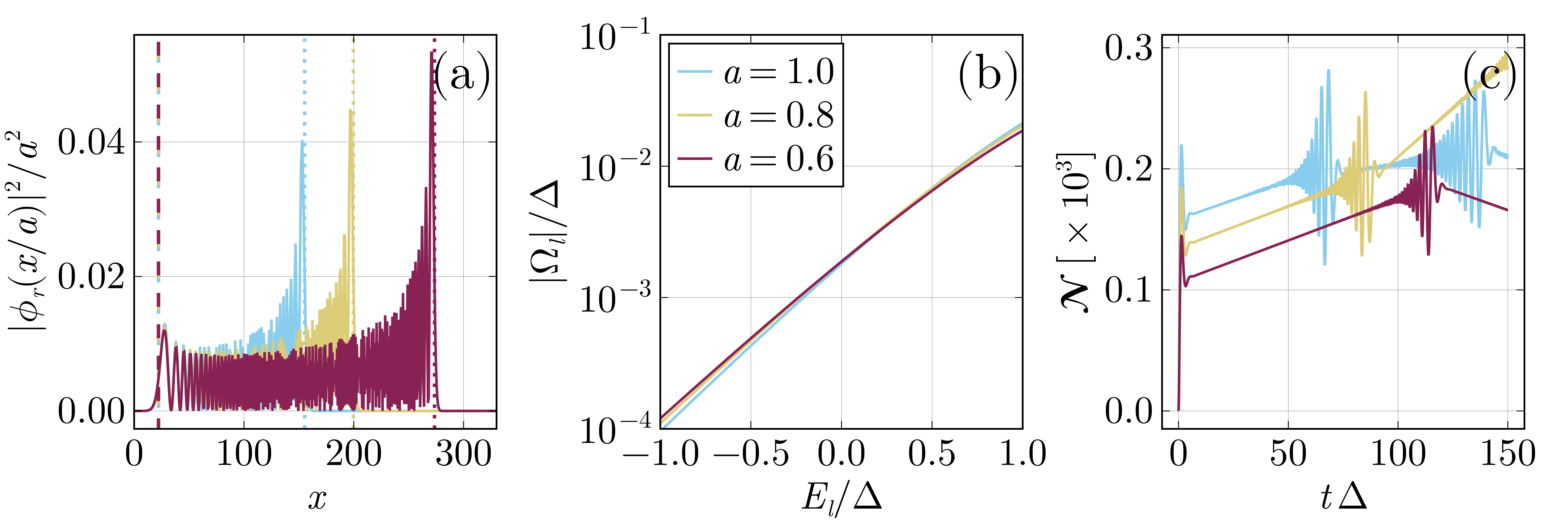}
    \caption{Behavior of different quantities under a scaling of the lattice spacing $a$. The parameters correspond to the parameters $p_{\chi,2}$ introduced in the main text, which are equivalent to $\Delta/\delta=11.0886$. The length of the system has been kept fixed at $X=360$. The low-energy physics is described by $\Delta$ and $\delta$ which are kept fixed while  scaling $a$. This is achieved by rescaling the external fields according to \cref{eq:hperp_scaling,eq:hlon_scaling}. (a) Single-bubble wave functions rescaled by $1/a$ to account for the increased spatial extent. The dashed/dotted vertical lines are the turning points in \cref{eq:turningPoints}. (b) Coupling between the FV and the $L$'th single-bubble eigenstate in \cref{eq:coupling_a}. (c) Total density of bubbles according to \cref{eq:totDen}.}
    \label{fig:scaling}
\end{figure*}
As a first verification of the scaling we plot in \pref{fig:scaling}{a} the wave-functions for three different values of $a$. 
As it was shown in the main text, the extent of the wave-function is described by the turning points 
\begin{equation}\label{eq:turningPoints}
x_r^{\pm}=l^\pm_r a=2a\frac{1\pm h_\perp}{\abs{h_\parallel} M}=\begin{cases}
			\frac{4}{\delta a}-x^-_r\\
            \frac{2 \Delta}{\delta}
		 \end{cases},
\end{equation}
where we have used the scalings derived above.
The lower turning point, which describes the classically expected bubble size is seen to be unaffected by the scaling,  while the large distance turning point describing the possible largest bubbles in the resonant eigenstate scales with $a^{-1}$. 
This growth leads to an overall extension of the wavefunction and the normalization of the wavefunction thus requires a decrease of its amplitude. 
As the extension of the wavefunction is almost proportional to $a^{-1}$, we can make up for the decrease in amplitude by also multiplying the wavefunction amplitude by $a^{-1}$ as done in \pref{fig:scaling}{a}. 
The good overlap between all three wavefunctions near $x^-_r$ shows that the physics due to the classically expected bubble size is maintained by our scaling of the external fields.    

Having derived the correct scaling of the external fields to have a well-defined continuum limit for the single-bubble EVP means that the energies also have a well-defined scaling limit. 
For the CBT theory, the two additional quantities are $\Omega_L$ and $\lambda_{L,L'}$. 

Considering first $\Omega_L$, we scale the wave-functions such that they have a well-defined continuum limit according to \cref{eq:wf_scaling} which leads to the appearance of the physical length $X$ in \cref{eq:FVbub_overlap}
\begin{equation}\begin{aligned}\label{eq:coupling_a}
    \Omega_L&=\frac{1}{2}\sqrt{X}\,M J\abs{h_\parallel}\sum_{n=1}^N h_\perp^n\frac{\phi_{L,n}}{\sqrt{a}}\\
    &=\frac{\delta}{4}\sqrt{X}\,a\sum_{n=1}^N h_\perp^n\frac{\phi_{L,n}}{\sqrt{a}}.
    \end{aligned}
\end{equation}
Taking the scaling limit for this expression, one finds the form
\begin{equation}
\label{eq:Omega_L_cont}\begin{aligned}
    \lim_{a\to0}\Omega_L&=\frac{\delta}{4}\sqrt{X}\int_0^Xdx \, (1-\Delta a)^n\psi_{L}(x)\\
    &=\frac{\delta}{4}\sqrt{X}\int_0^Xdx  \,e^{-\Delta x}\psi_{L}(x),
    \end{aligned}
\end{equation}
which shows that the coupling scales to a well-defined value in the continuum limit. 
In \eqref{eq:Omega_L_cont}, the characteristic length $1/\Delta$ physically corresponds to the thickness of the bubble walls. 
As seen in \pref{fig:scaling}{b}, this low-energy (near $E=0$) behavior of $\Omega_L$ is only weakly perturbed indicating that the low energy behavior is largely preserved when the external fields are appropriately scaled.

Consider now $\lambda_{L,L'}$ computed in \cref{eq:lambda}. 
As for $\Omega_L$ we first explicitly include the $\sqrt{a}$ that gives the wave function a continuum limit and then rewrite the constants into a combination of $\delta$ and $a$
\begin{equation}
\begin{aligned}
     \lambda_{L,L'}&=-J\abs{h_\parallel} 2M\sum_{n=1}^Nna\frac{\phi_{L,n}\phi_{L',n}}{a}\\
     &=-\delta a\sum_{n=1}^Nna\frac{\phi_{L,n}\phi_{L',n}}{a},
\end{aligned}
\end{equation}
In the scaling limit, this takes the form
\begin{equation}
\begin{aligned}
     \lim_{a\to0}\lambda_{L,L'}&=-\delta \int_0^X dx \,x\,\psi_{L}(x)\psi_{L'}(x),
\end{aligned}
\end{equation}
which also is well-defined in the continuum limit. 

Having shown that all the quantities needed for the CBT have a well-defined continuum limit, we
now consider the scaling of the total density in \cref{eq:totDen}
\begin{equation}
    \mathcal{N}=\frac{2 t}{N}\sum_L \abs{\Omega_L}^2\frac{\sin^2\left(E_L t/2\right)}{E_L^2 t/2}.
\end{equation}
Using the density of states $\rho(E)=\sum_L \delta(E-E_L)$ we can replace the sum with an integral over energy
\begin{equation}\label{eq:totDen_Scaling1}
    \mathcal{N}=\frac{2 t}{N}\int dE\,\rho(E) \abs{\Omega_E}^2\frac{\sin^2\left(E t/2\right)}{E^2 t/2}.
\end{equation}
As discussed in the main text, near the resonant state $E_L$ is linearly spaced resulting in a constant density of states $\rho(E)=1/a \delta$. 
Additionally, we have the property 
\begin{equation}
    \lim_{t\to\infty}\frac{\sin^2\left(E t/2\right)}{E^2 t/2}=\pi \delta(E).
\end{equation}
Inserting this into \cref{eq:totDen_Scaling1} 
\begin{equation}\label{eq:N_FGR_adep}
    \mathcal{N}=\Gamma t,
\end{equation}
where $\Gamma$ is the FVD decay rate given by
\begin{equation}\label{eq:Gamma_discr}
    \Gamma =\frac{2 \pi \abs{\Omega_{E=0}}^2}{Na \delta}
\end{equation}
Using \cref{eq:Omega_L_cont} we find the continuum limit of $\Gamma$ to take the form
\cref{eq:totDen_Scaling1} 
\begin{equation}
    \Gamma=\frac{\pi \delta}{8} \abs{\int dx e^{-\Delta x} \psi_r(x)}^2,
\end{equation}
showing that in the continuum limit the linear behavior is dominating and has a well-defined slope. 
To verify this, $\mathcal{N}$ has been plotted in \pref{fig:scaling}{c} for three different lattice spacings.
It is observed that the slope is not affected by the scaling whereas the initial non-linear behavior is suppressed and the time scale for the Bloch oscillation extended, such that in the scaling limit only the linear slope remains. 

\section{Comparison between sudden quench and slow ramp \label{app:ramp}}
In this Appendix we present MPS calculations for the same evolution considered in the main text, but with the sudden quench replaced by a linear ramp of the longitudinal field $h_\parallel$ defined by
\begin{equation}
    R_{\tau}(t) = \begin{cases}
        \frac{th_\parallel}{\tau}  &\text{if } 0\le t <\tau, \\
        h_\parallel & \text{if } t\ge\tau
    \end{cases}
    \label{eq:ramp_shape}
\end{equation}
with a characteristic ramp time $\tau$.
In \pref{fig:ramp}{a}, we show a series of plots of the time-evolution of the magnetization $m(t)$ for different values of the ramp time $\tau$.

As the ramp time $\tau$ is increased, the short-time downward jump in the magnetization that is visible in the dark curves disappears and is replaced by the smooth decrease with an approximately parabolic law that is visible in the brighter curves. This confirms the fact that the excitation of high-energy bubbles at short times and the consequent drop in magnetization are a consequence of the abrupt quench and are suppressed when the ramp is sufficiently slow.
\begin{figure*}
    \centering
    \includegraphics[width=0.85\linewidth]{mag_ramp_plot.png}
    \caption{MPS calculation of the magnetization $m(t)$ for the same parameters set as $X_1$. As a key difference from the main text, the sudden quench at $t=0$ is replaced by a smooth ramp of $h_\parallel$ of the form \cref{eq:ramp_shape} with different ramp times $\tau$. The ramp time is indicated by the color coding which can be read of from shared colorbar. (a) Time evolution of the magnetization. (b) The second derivative of $m(t)$, with a shifted time argument, $t_m$, such that the first minima after the ramp is at 0 on the x-axis. The value of $t_m$ is indicated by the dashed vertical lines in (a), where the intersection points are marked by a circle which are connected by a dotted black line serving as a visual guide. The red solid line in (b) is showing the approximation in \cref{eq:mdd_approx}, whereas the red dashed line is showing \cref{eq:tdep_speed}.  
    \label{fig:ramp}}
\end{figure*}

Extracting the universal physics from these curves is complicated by the peculiarity of the one-dimensional geometry that the high-energy bubbles that are created in the short-time transient eventually expand under the effect of the repulsive interactions between the fermions. Given the difficulty of removing this contribution in an accurate way,
there is no straightforward way of superimposing the late-time curves of the magnetization $m(t)$ for different values of the ramp time $\tau$ and a more sophisticated analysis is required. 

The key idea is that the FVD contribution to the magnetization has a super-linear growth whereas the expansion of the high-energy bubbles created at early times gives an approximately linear contribution. As a result, one can separate the two effects by considering the second derivative of the magnetization in time $\ddot{m}(t)$ which is only sensitive to the FVD.  
The connection between $\ddot{m}(t)$ and the FVD rate $\Gamma$ can be quantified using a classical picture. 
First, we write the magnetization as the sum of its initial value and the contribution from a density $n_l$ of bubbles with size $l$.
Summing over all bubble sizes, one has the approximate expression 
\begin{equation}\label{eq:clas_disc_mag}
    m(t)\approx M-2M \sum_{l=0}^{\infty} l n_l(t).
\end{equation}
Within a continuum approximation for $l$, we can write a transport equation for $n_l(t)$ which contains a flux term where bubbles expand with a speed $v(t)$ and a source term for the classical resonant bubble size $l_r$
\begin{equation}\label{eq:cont_nl}
    \dot{n}_l(t)=-v(t) \frac{d n_l(t)}{d l}+\Gamma \delta(l-l_r).
\end{equation}
From the semi-classical result in \cref{eq:blochOsc} we have that $v(t)$ is given by
\begin{equation}\label{eq:speed}
    v(t)=\frac{d \,r(t)}{dt}=\frac{8 J^2 h_\perp \sin\left(2 h_\parallel M t\right)}{\epsilon\left(2h_\parallel J M t\right)},
\end{equation}
meaning that the speed of the bubble depends on when it was created. As our simple estimate here does not allow for giving different bubbles different speeds the most consistent approximation is to assume the speed to be constant for all bubbles. From \pref{fig:evo}{g} we know this is a reasonable approximation already at early times and therefore use a constant velocity $v$, determined from the constant slope of \pref{fig:evo}{g}. 
Notice that neither \cref{eq:cont_nl,eq:speed} contain the ramp time and are therefore only valid after the ramp. Importantly this means that the predictions based on these results will be independent of how the quench is done.
Combining \cref{eq:cont_nl} with the time derivative of the continuum version of \cref{eq:clas_disc_mag} one finds
\begin{equation}\label{eq:mdot}
\begin{aligned}
    \dot{m}(t)&=-2M\int_0^\infty dl \;l\;\left(-v \frac{d n(l,t)}{dl}+\Gamma \delta(l-l_r)\right)\\
    &=-2M\left(v\mathcal{N}(t) + \Gamma l_r\right),
\end{aligned}
\end{equation}
which only holds for finite times where $n(\infty,t)=0$. Lastly, taking the time derivative of \cref{eq:mdot} the second derivative of the magnetization takes the form
\begin{equation}\label{eq:mdd_approx}
    \ddot{m}(t)=-2Mv\dot{\mathcal{N}}(t)=-2Mv\Gamma,
\end{equation}
which follows from \cref{eq:N_FGR_adep}.
Computing $\Gamma$  from \cref{eq:Gamma_discr}, as done in \pref{fig:evo}{e}, in \pref{fig:ramp}{b} we have plotted this semi-classical approximation together with the numerically computed second derivatives.
Each curve is shifted by the time $t_m$, chosen so that the first minimum after the ramp occurs at $0$ on the x-axis. 
In \pref{fig:ramp}{a} one sees that at this time the magnetizations are all similar. The main uncertainty arises from the high-energy oscillations in \pref{fig:ramp}{b}, which leads to shifts in the determination of $t_m$. 
Looking at \pref{fig:ramp}{b} and focusing on the positive part of the  x-axis, one observes that the shorter quenches contain high-energy excitations resulting in fast oscillations but the slow dynamics set by the FVD dynamics collapses for all curves.
The fact that the slow dynamics, after a certain decay of the magnetization, is independent of the ramp shows that the FVD dynamics is itself independent of how the ramp was made. The analysis thus shows that the conclusions in the main section can immediately be extended beyond the instantaneous quench protocol.

The crude approximation in \cref{eq:mdd_approx} predicts a constant value for $\ddot{m}(t)$, which is seen to be a very good quantitative measure for the minimum of $\ddot{m}$. 
There are two simple mechanisms that can explain why $\ddot{m}$ does not remain constant.  
One is that the speed of the bubbles is not constant but time-dependent as already seen in the semi-classical description in \cref{eq:speed}. 
To properly describe this one would need to use something equivalent to a time-dependent version of the CBT. Extending the CBT to be time-dependent does not pose any fundamental challenges but goes beyond the scope of our article. However, to show that this effect is indeed relevant we can ignore the inconsistency of the approximation and replace the constant velocity in \cref{eq:cont_nl} with the semi-classical result in \cref{eq:speed}. 
This leads to an approximation for $\ddot{m}(t)$ of the form
\begin{equation}\label{eq:tdep_speed}
    \ddot{m}(t)=-2Mv(t)\dot{\mathcal{N}}(t)-2M\dot{v}(t)\mathcal{N}(t)\approx-2Mv(t)\Gamma
\end{equation}
where the second term can be ignored due to the slow variation of the speed (see \pref{fig:evo}{e}).
To set the "initial" time we can repeat the numerical procedure and shift the minimum to zero on the x-axis using the shift
\begin{equation}
    t_{m,sc}=\frac{\cos^{-1}\left(h_\perp\right)}{2JMh_\parallel},
\end{equation}
which follows from the maximum of \cref{eq:speed}. 
Including this effect captures some of the decrease in magnitude of $\ddot{m}(t)$ but a time-dependent CBT would be necessary to capture both the high-energy oscillations and the fact that the consistent description would correlate the speed with the time at which the bubble is created.  

The other effect that could explain the decrease of $\abs{\ddot{m}}(t)$ observed in \pref{fig:ramp}{b} is the increasingly important role of interactions. 
A further reduction in the magnitude of $\ddot{m}(t)$ beyond the reduction predicted via the approximated time-dependent speed is in fact observed in the Figure. This deviation is likely connected to the gradual increase of interactions. As already discussed in \cref{sec:XB}, these interactions appear to be repulsive and therefore tend to slow down the bubbles, further decreasing the magnitude of $\ddot{m}(t)$. 
As the effect of interactions would not be included within a time-dependent CBT, its deviation from the numerical results would help disentangle the effects of interactions and the time-dependent speed.

\bibliography{FVD_bib} 
\end{document}